%% file: New_Auction_Assignment_ARXIV.tex
\input TEXSHOP_macros_new.tex

  
\def\texshopbox#1{\boxtext{462pt}{\vskip-1.5pc\pshade{\vskip-1.0pc#1\vskip-2.0pc}}}
\def\texshopboxnt#1{\boxtextnt{462pt}{\vskip-1.5pc\pshade{\vskip-1.0pc#1\vskip-2.0pc}}}
\def\texshopboxnb#1{\boxtextnb{462pt}{\vskip-1.5pc\pshade{\vskip-1.0pc#1\vskip-2.0pc}}}










%

%


\input epsf




\long\def\fig#1#2#3{\vbox{\vskip1pc\vskip#1
\prevdepth=12pt \baselineskip=12pt
\vskip1pc
\hbox to\hsize{\hfill\vtop{\hsize=30pc\noindent{\eightbf Figure #2\ }
{\eightpoint#3}}\hfill}}}

\def\show#1{}
\def\frac#1#2{{#1\over #2}}
\def\jstar{J^{\raise0.04pt\hbox{\sevenpoint *}} }

\rightheadline{\botmark}

\pageno=1


\pn {\bf October 2023}\hfill{\bf Arizona State University/SCAI Report}
\bigskip \bigskip

\bigskip\bigskip\bigskip

\def\longpapertitle#1#2#3{{\bf \centerline{\helbigb
{#1}}}\medskip{\bf \centerline{\helbigb
{#2}}}\medskip{\centerline{
by}}\medskip{\bf \centerline{
{#3}}}\bigskip}

\longpapertitle{New Auction Algorithms for the}{Assignment Problem and Extensions \footnote{\dag}{\ninepoint Many thanks are due to Yuchao Li for extensive helpful comments.}}
{{Dimitri Bertsekas\ \footnote{\ddag}{\ninepoint Fulton Professor of Computational Decision Making, School of Computing and Augmented Intelligence, Arizona State University, Tempe, AZ.}}}

\centerline{\bf Abstract}

We consider the classical linear assignment problem, and we introduce new auction algorithms for its optimal and suboptimal solution. The algorithms are founded on duality theory, and are related to  ideas of competitive bidding by persons for objects and the attendant market equilibrium, which underlie real-life auction processes. We distinguish between two fundamentally different types of bidding mechanisms: {\it aggressive} and {\it cooperative\/}. Mathematically, aggressive bidding relies on a notion of approximate coordinate descent in dual space, an $\e$-complementary slackness condition to regulate the amount of descent approximation, and the idea of $\e$-scaling to resolve efficiently the price wars that occur naturally as multiple bidders compete for a smaller number of valuable objects. Cooperative bidding avoids price wars through detection and cooperative resolution of any competitive impasse that involves a group of persons. 

We discuss the relations between the aggressive and the cooperative bidding approaches, we derive new algorithms and variations that combine ideas from both of them, and we also make connections with other primal-dual methods, including the Hungarian method. Furthermore, our discussion points the way to algorithmic extensions that apply more broadly to network optimization, including shortest path, max-flow, transportation, and minimum cost flow problems with both linear and convex cost functions.

\vfill\eject

\section{Introduction}
\vskip-1pc

\pn  In this paper, we discuss auction algorithms for solving numerically the classical assignment (aka weighted bipartite matching) problem, where there are $n$ persons, denoted by $i=1,\ldots,n$, and $n$ objects, denoted by $j=1,\ldots,n$, which we have to match on a one-to-one basis. Each person $i$ may be matched to any object $j$ within a given subset $A(i)\subset \{1,\ldots,n\}$. By  a {\it complete assignment} we mean a set of person-object pairs $(1, j_1),\ldots, (n, j_n)$, such that $j_i\in A(i)$ for all $i=1\ldots,n$, while the objects $j_i$ are all distinct. There is a known value $a_{ij}$ for matching person $i$ with object $j\in A(i)$, and we want to find a complete assignment that maximizes the total value 
$$\sum_{i=1}^ n a_{ij_i}.$$

The assignment problem has received a lot of attention since the 1950s. It arises in many practical settings, the most obvious ones being resource allocation problems, such as assigning personnel to jobs, resources to tasks, and related contexts, such as scheduling and data association.  The assignment problem also appears often as a subproblem in various methods for solving more complex problems. 

Recent applications of the assignment problem include:

\nitem{(a)} {\it Optimal transport} (arising in cosmology among others; see e.g., Brenier et al.\ [BFH03], Frisch and Sobolevskii [FrS06], Lavaux [Lav08], Villani [Vil09], [Vil21],  Santambrogio [San15], Galichon [Gal16], Metivier et al.\ [MBM19], Schmitzer [Sch16], [Sch19],  Walsh and Dieci [WaD17], [WaD19], Peyre and Cuturi [PeC19], Levy, Mohayaee, and von Hausegger [LMH21], Merigot and Thibert [MeT21]).

\nitem{(b)} {\it Graph similarity problems} (arising in computational biology among others; see e.g., Kollias at al.\ [KSS14], Erciyes [Erc15]).

\nitem{(c)} {\it Graph neural networks} (see e.g., Zhou et al.\ [ZCH20],   Aironi, Cornell, and Squartini [ACS22], Nurlanov, Schmidt, and Bernard [NSB23]),

\nitem{(d)} {\it Combinatorial auctions} (see e.g., Parkes and Ungar [PaU99], De Vries and Vohra [DeV03]).

\nitem{(e)} {\it Computational physics} (see e.g., Kosowsky and Yuille [KoY94], Jacobs, Merkurjev, and Esedoglu [JME18], Bertozzi and Merkurjev [BeM19], [Mer20]).

\nitem{(f)} {\it A variety of dynamic task allocation, scheduling, multiagent, and multi-robot problems} (see e.g.,  Bayati et al.\ [BPS07], Bayati, Shah, and Sharma [BSS08], Choi, Brunet, and How [CBH09], Liu and Shell [LiS13], Luo, Chakraborty, and Sycara  [LCS14], Morgan et al.\ [MSC16], Tang et al.\ [TZG18], Duan et al.\ [DLT19], Huang, Zhang, and Xiao [HZX19], Luzak et al.\ [LuM20], [LGO20], Otte, Kuhlman, and Sofge [OKS20], Aziz et al.\ [APP22], Wang et al.\ [WMW22], Garces et al.\ [GBG23], Li et al.\ [LZX23], and Wang, Li, and Yao [WLY23]). 
\smskip

The assignment problem is also of great theoretical significance because, despite its simplicity, it embodies a fundamental linear programming structure. In particular, the important single commodity linear cost network flow problem can be reduced to an assignment problem by means of a simple reformulation. Thus, any method for solving the assignment problem can be
generalized to solve the linear network flow problem, and in fact this approach is particularly helpful in understanding the extensions of auction algorithms to network flow problems that are more general than assignment. Detailed discussions can be found in the author's network optimization textbooks [Ber91a], [Ber98], and are very relevant to the research directions presented in this paper.

\subsubsection{Duality Theory for the Assignment Problem}

\pn To develop an intuitive understanding of  auction algorithms, it is helpful to introduce an economic equilibrium problem that turns out to be equivalent to the assignment problem. Let us consider the possibility of matching the $n$ objects with the $n$ persons through a market mechanism, viewing each person as an economic agent acting in his/her own best interest. Suppose that object $j$ has a price $p_j$ and that the person who acquires the object must pay the price $p_j$. Then the (net) profit of object $j$ for person $i$ is $a_{ij}-p_j$ and each person $i$ would logically want to be assigned to a maximal profit object $j_i\in A(i)$, i.e., one satisfying
$$a_{ij_i}-p_{j_i} = \max_{j\in A(i)} \{a_{ij}-p_j\}.$$
A set of prices $p=(p_1,\ldots,p_n)$ and a set of assigned pairs ${\cal A}=\big\{(i_1, j_1),\ldots, (i_k, j_k)\big\}$ where each assigned person satisfies the preceding condition, i.e.,
$$a_{i_mj_m}-p_{j_m} = \max_{j\in A(i_m)} \{a_{i_mj}-p_j\},\qquad \hbox{for all assigned pairs }(i_m,j_m)\in{\cal A},\xdef\compslackfinal{\lab}\eqnum\show{oneo}$$
are said to satisfy {\it complementary slackness} (CS for short).  When CS holds, for a set of prices $p$ and a complete assignment ${\cal A}$ (i.e., one where $k=n$), we have a form of economic equilibrium whereby each person is assigned to an object that offers maximum profit, and has no incentive to switch to a different object.

A fundamental duality theorem states that a complete assignment that satisfies the CS condition \compslackfinal\ together with some set of prices, is optimal, i.e., it offers maximum total value. Moreover, the corresponding set of prices solves an associated dual optimization problem, which is to minimize over $p=(p_1,\ldots,p_n)$ the dual cost function
$$\sum_{i=1}^n \p_i+\sum_{j=1}^np_j,\xdef\dualcost{\lab}\eqnum\show{oneo}$$
where $\p_i$ is the maximum profit that is attainable for person $i$ under the set of prices $p$:
$$\p_i=\max_{j\in A(i)}\{a_{ij}-p_j\},\qquad i=1,\ldots,n.\xdef\profit{\lab}\eqnum\show{oneo}$$
Mathematically, we can view this as a consequence of the celebrated duality theorem of linear programming, whereby the assignment optimization is viewed as the primal problem and the minimization of the cost \dualcost-\profit\ is the dual problem.\footnote{\dag}{\ninepoint The proof is very simple. For any set of prices $(p_1,\ldots,p_n)$ and any complete assignment $(1, j_1),\ldots, (n, j_n)$, using the definition \profit\ of the profit $\p_i$, we have
$$\sum_{i=1}^n \p_i+\sum_{j=1}^np_j=\sum_{i=1}^n \max_{j\in A(i)}\{a_{ij}-p_j\}+\sum_{j=1}^np_j\ge\sum_{i=1}^n \{a_{ij_i}-p_{j_i}\}+\sum_{j=1}^np_j= \sum_{i=1}^ n a_{ij_i}.$$
Under the CS condition \compslackfinal, equality holds in the above relation. Thus when CS is satisfied, $(p_1,\ldots,p_n)$ attains the minimum of the dual cost on the left side above, while $(1, j_1),\ldots, (n, j_n)$ attains the maximum of the right side.}

\subsubsection{Algorithms for Solving the Assignment Problem}

\pn There are several iterative algorithms for the solution of the assignment problem, which are described in detail in several sources, including the linear programming textbook by Bertsimas and Tsitsiklis [BeT97], the network optimization books by Bertsekas [Ber91a], [Ber98], and Burkard, Dell'Amico, and Martello [BDM12], and the extensive surveys by Ahuja, Magnanti, and Orlin [AMO88], [AMO89], and Burkard and Cela [BuC99], among others. In particular, there are:

\nitem{(a)} {\it Primal simplex methods\/}, which start with some feasible assignment (a primal solution) and iteratively increase the value of the assignment by using the mechanism of the simplex method, suitably adapted to take advantage of the underlying graph structure. 

\nitem{(b)} {\it Dual cost improvement methods\/}, which include Kuhn's  Hungarian method [Kuh55], and the relaxation algorithms by Bertsekas [Ber81], [Ber85], and Bertsekas and Tseng [BeT88]. These methods start from a dual solution (a set of object prices) and iteratively modify the prices along dual descent directions, thus generating a cost improving sequence of dual solutions. 

A third and distinct class of iterative methods for the assignment problem is {\it auction algorithms\/}, the subject of this paper. These methods resemble real-life auctions and can be loosely interpreted as approximate coordinate descent methods for solving the dual problem. The approximation is controlled by a parameter $\e>0$, which may be reduced in the course of the algorithm. Auction algorithms differ from primal methods and dual methods in a fundamental way: they may deteriorate both the primal and the dual objectives at any one iteration by an amount that depends on $\e$. Still, with appropriate implementation and control of the size of $\e$, they find an optimal primal and dual solution pair.

\subsubsection{Aims and Contributions of the Paper}

\pn The present paper focuses on three related types of auction algorithms, {\it conservative}, {\it aggressive}, and {\it cooperative}, which aim to find a set of prices and a complete assignment that attain the market equilibrium noted earlier, and hence solve the corresponding dual and primal problems.
The aggressive auction algorithm was first proposed by the author in the paper [Ber79], and was followed by a proposal of a cooperative auction algorithm in the paper [Ber81]. The conservative auction algorithm, which is a limiting form of the aggressive auction algorithm, was also discussed in these papers, and in fact it was suggested as an effective initialization of some of the cooperative algorithms of the paper [Ber81], despite the fact that in general it does not guarantee convergence to an optimal assignment.\footnote{\dag}{\ninepoint The term ``naive auction" was used instead of ``conservative auction" in these and other subsequent works. We will avoid the term ``naive" in this paper: it is somewhat misleading because conservative auction embodies interesting ideas, and is useful both conceptually and practically, despite the fact that it does not guarantee convergence to an optimal assignment. The paper [Ber81] also proposed and tested a two-phase algorithm, whereby conservative/naive auction was used in the first phase to initialize a Hungarian algorithm used in the second phase.  The code of Jonker and Volgenant [JoV87], often referred to as the ``JV code," is very similar. It uses the conservative auction algorithm to initialize a Hungarian-like sequential shortest path method, but starts conservative auction with the classical choice for initial prices: $p_j$ is set to $\min_{i}a_{ij}$, rather than $p_j=0$, the author's choice in the code of [Ber81] (in fact the authors of [JoV87] developed their code working from a printout of the author's 1981 code). The JV code has been used widely, as it clearly performs better than codes based on the classical Hungarian method, thanks to its conservative auction initialization. On the other hand, aggressive auction codes seem to outperform the JV code, and other Hungarian-related codes, for many types of problems, although assessments differ on this issue; see e.g., Bertsekas and Eckstein [BeE88], Casta\~ non [Cas93], Zaki [Zak95], Malkoff [Mal97]. Aggressive auction codes also seem to outperform codes that are inspired by preflow-push ideas (whose mechanism can be viewed as mathematically equivalent to the one of the aggressive auction algorithm); see the papers by Bertsekas [Ber93], Naparstek and Leshem [NaL16], Alfaro et al.\ [APV22], and the textbook [Ber98] (Section 7.3.3).}

The distinction between the conservative and aggressive auction algorithms can be described in terms of a critical parameter $\e$ that characterizes the ``intensity" of competition between the persons for the objects: in conservative auction $\e=0$, while in aggressive auction $\e>0$. The cooperative auction algorithm, as given in [Ber81], uses $\e=0$, so it has a conservative character. The present paper extends substantially the cooperative auction framework by allowing $\e>0$ and by integrating the three different types of auction into a single method, aiming to combine their best characteristics.  In particular, the extension to the case where $\e>0$ involves qualitatively significant changes in the algorithm's character, and appears to be substantially faster for many problems. The new ideas of this paper also point the way towards extensions to network optimization problems that are more general than assignment. 

We  first review in Section 2 some of the known ideas relating to conservative and aggressive auctions, and the principal challenges that they face due to what we will call {\it competitive impasses} and {\it price wars\/}. 
In Section 3, we  propose a new cooperative auction algorithm,  which aims to provide a mechanism for addressing price wars. We discuss several variations, including the {\it expanding coalitions variant} of cooperative auction, which provides a conceptual vehicle for bridging the ideas of auction and Hungarian methods. The algorithm is structured so that it can combine harmoniously conservative, aggressive, and cooperative auction ideas. A combination of this type was given in  the paper [Ber81] for the special case where $\e=0$. We provide a similar combination, but one where $\e>0$. In Section 4, we discuss additional variations of the algorithms of Sections 2 and 3, as well as the role of $\e$-scaling within the broader cooperative auction framework of the paper.

In the present paper we will focus on the algorithmic ideas underlying auction algorithms for the assignment problem, particularly the new cooperative versions. In a future report, we will provide results of computational experimentation and describe how our auction ideas can be extended to other linear network flow problems, such as shortest path, max-flow, transportation, and transshipment problems. We will also extend our algorithms of Sections 3 and 4 to single commodity network flow problems with separable convex cost functions, building on auction algorithmic ideas presented in the papers by Bertsekas, Polymenakos, and Tseng [BPT97], [BPT98], and discussed in more detail in Chapter 9 of the book [Ber98].
 
\vskip-1.5pc

\section{Conservative and Aggressive Auctions}
\vskip-0.5pc

\xdef\figconservativepricechange{\figr}\figrnum\show{myfigure}

\pn Let us first establish some terminology. In what follows, by an {\it assignment} we mean a set of person-object pairs $(i_1, j_1),\ldots, (i_k, j_k)$, such that $j_1\in A(i_1),\ldots,j_k\in A(i_k)$, while $i_1,\ldots,i_k$ are distinct persons and $j_1,\ldots,j_k$ are distinct objects. If $k=n$ the assignment is called {\it complete\/}, and if $k<n$ the assignment is called {\it partial} (or {\it incomplete\/}). The empty assignment, where there are no assigned persons or objects, is also considered to be a partial assignment. Generally, assignments (complete, incomplete, or empty) will be denoted by ${\cal A}$.  We assume throughout that there exists at least one complete assignment for our given problem. Also for simplicity in describing algorithms, and without loss of generality, we assume that $A(i)$, the set of objects to which person $i$ can be assigned, contains at least two elements.

A common characteristic of all auction algorithms is that they maintain at all times a partial assignment ${\cal A}$ and a set of object prices $p=(p_1,\ldots,p_n)$, which satisfy an approximate form of the  CS condition \compslackfinal\ that involves a parameter $\e\ge0$. The partial assignment grows progressively to become a complete assignment, at which time the auction algorithm terminates. 

The central mechanism of an auction algorithm is a bid by an unassigned person $i$ for his/her ``best" object $j_i$ (the one that maximizes the person's profit):
$$a_{ij_i}-p_{j_i}=\max_{j\in A(i)} \{a_{ij}-p_j\}.$$
In particular, person $i$ bids for $j_i$ by raising its price from $p_{j_i}$ to $\ol p_{j_i}$ given by 
$$\ol p_{j_i}=\e+\hbox{the price level that makes the profit of $j_i$ equal to the second best profit }
\max_{j\in A(i),\, j\ne j_i} \{a_{ij}-p_j\}.$$

Depending on whether $\e=0$ or $\e>0$, the auction is called {\it conservative} or {\it aggressive\/}, respectively. Thus in an aggressive auction the object prices are raised by larger increments. 
In this section we will review these two different types of auction and their properties. For detailed discussions, which include additional topics, such as parallel and asynchronous distributed implementations, we refer to the textbooks [BeT89], [Ber91a], and [Ber98], and the tutorial papers [Ber90] and [Ber92]. No new research is presented in this section. 

\vskip-0.5pc
\subsection{Conservative Auction}
\vskip-0.5pc
\pn As in real-life auctions, a person needs to balance two competing considerations when  determining a proper bid size: a high bid for his/her preferred object discourages bids of other persons for that object, but also diminishes his/her profit upon acquiring that object. Thus it makes sense for a person $i$ to maximize the bid for a preferred object $j_i$ subject to the constraint that this object continues to offer maximum profit, i.e., to raise the price of $j_i$ to
$$\ol p_{j_i}=a_{ij_i}-w_i,$$
where $w_i$ is the ``second best" profit,
$$w_i= \max_{j\in A(i),\, j\ne j_i} \{a_{ij}-p_j\},$$
thereby bringing the profit 
$$ a_{ij_i}-\ol p_{j_i},$$
of the best object $j_i$ to the level of the profit $w_i$ of the second best object; see Fig.\ \figconservativepricechange.  We view this auction mechanism as conservative because when selecting a bid, person $i$ takes no risk, in the sense that he/she will never end up with a non-maximum profit object.\footnote{\dag}{\ninepoint Price rises below the maximum level $a_{ij_i}-w_i$ also have this property, but larger price rises tend to accelerate the termination of the auction, and are therefore better suited for our algorithmic purposes.}

\topinsert
\centerline{\hskip0pc\epsfxsize = 3.3in \epsfbox{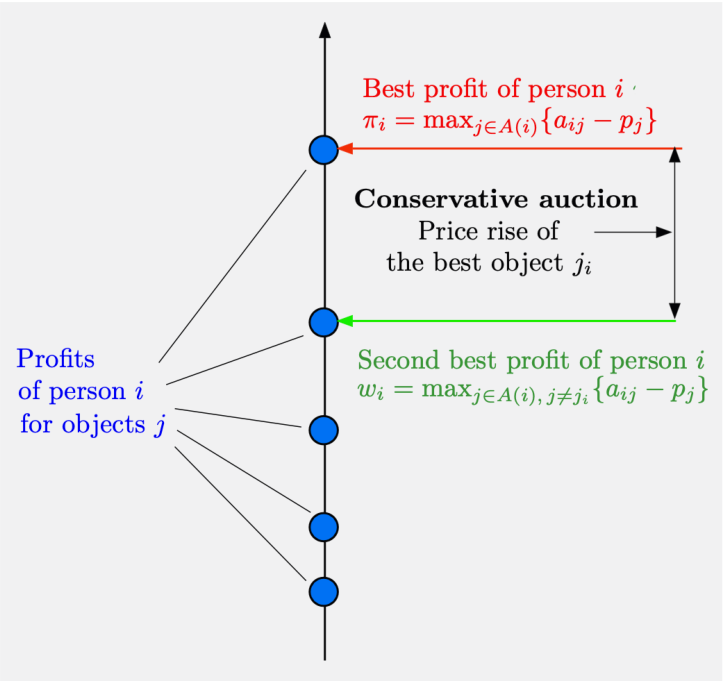}}
\vskip-0.5pc
\hskip-3pc\fig{-1pc}{\figconservativepricechange} {Illustration of the price rise of the best object $j_i$ of an unassigned person $i$ in the conservative auction algorithm. The price of $j_i$ is increased by $\p_i-w_i$, while the profit of $j_i$ is made equal to $w_i$.}\endinsert

Let us describe the conservative auction  algorithm more precisely. The algorithm proceeds in iterations and  throughout its operation, maintains a set of prices $p=(p_1,\ldots,p_n)$ and a partial assignment ${\cal A}$  where each assigned person is assigned to a maximal profit object, i.e., the CS condition \compslackfinal\ is satisfied. It terminates when following an iteration, the assignment obtained is complete. The algorithm starts with any set of prices and partial assignment that satisfy CS; for example it may start with an arbitrary set of prices and the empty assignment. Given the current set of object prices $p$ and partial assignment ${\cal A}$, a conservative auction iteration generates a new set of prices and a new assignment as follows.

\texshopbox{\pn {\bf Conservative Auction Iteration}
\smskip
\pn We select an unassigned person $i$ and an object $j_i$ that offers maximum profit for $i$ under the given prices,
$$a_{ij_i}-p_{j_i} = \max_{j\in A(i)} \{a_{ij}-p_j\}.\xdef\bestprofit{\lab}\eqnum\show{oneo}$$
 We set the price of $j_i$ to
$$\ol p_{j_i}=a_{ij_i}-w_i,\eqnum\show{oneo}$$
where $w_i$ is the ``second best" profit,
$$w_i= \max_{j\in A(i),\, j\ne j_i} \{a_{ij}-p_j\}.\xdef\secondbest{\lab}\eqnum\show{oneo}$$
Finally, we add to the assignment ${\cal A}$ the pair $(i,j_i)$, and if $j_i$ was assigned to some other person $\tl i$, we remove from ${\cal A}$ the pair $(\tl i,j_i)$, thus forming a new assignment $\bar{\cal A}$.
}

\xdef\figthreebythree{\figr}\figrnum\show{myfigure}

\topinsert
\centerline{\hskip0pc\epsfxsize = 4.0in \epsfbox{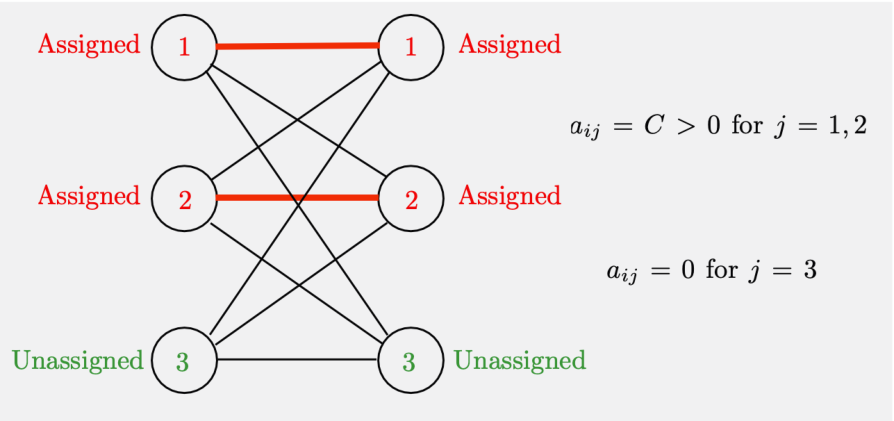}}
\vskip-1pc
\vskip0pt
\eightpoint
\def\tablerule{\noalign{\hrule}}
$$\vbox{\offinterlineskip
\hrule
\halign{\vrule\hfill \ #\ \hfill &\vrule\hfill \ #\ \hfill 
&\vrule\hfill \ #\ \hfill &\vrule\hfill \ #\ \hfill 
&\vrule\hfill \ #\ \hfill &\vrule\hfill 
\ #\ \hfill \vrule\cr
&&\cr
{\bf At Start of\lower2ex\hbox{\ }\raise4ex\hbox{\ }}&\hbox{\bf
Object}&\hbox{\bf Assigned}&\hbox{\bf Bidding}&\hbox{\bf Best}&\hbox{\bf Bidding}\cr
{\bf Iteration \#\lower2ex\hbox{\ }}&\hbox{\bf  Prices}&\hbox{\bf  Pairs} &\hbox{\bf Person}&\hbox{\bf Object} &\hbox{\bf Increment}\cr
\tablerule\cr &&\cr
{1\lower2ex\hbox{\ 
}\raise4ex\hbox{\ }\hbox{\ }}&$(0,0,0)$&$(1,1),(2,2)$&3&2&0\cr 
{2\lower2ex\hbox{\
}\raise2ex\hbox{\ }\hbox{\ }}&$(0,0,0)$&$(1,1),(3,2)$&2&2&0\cr 
{3\lower2ex\hbox{\
}\raise2ex\hbox{\ }\hbox{\ }}&$(0,0,0)$&$(1,1),(2,2)$&3&2&0\cr 
&&\cr \tablerule\cr}}$$
\fig{-1pc}{\figthreebythree} {Illustration of how the conservative auction algorithm may never terminate for a $3\times 3$ assignment problem. Here objects 1 and 2 have value $C>0$ for all persons, and object 3 has value 0 for all persons. The algorithm starts from the initial prices $p=(0,0,0)$ and the partial assignment $\big\{(1,1),(2,2)\big\}$. There is a competitive impasse involving persons 1, 2, and 3, and objects 1 and 2. The algorithm cycles as persons 2 and 3 alternately bid for object 2 (or object 1) without changing its price because they prefer equally object 1 and object 2. 
}\endinsert


It can be seen that the conservative auction algorithm maintains the CS condition \compslackfinal\ throughout its operation, and generates a sequence of partial assignments whose cardinalities are not decreasing, so if it terminates, the complete assignment obtained at termination is optimal, while the corresponding final prices are an optimal solution to the dual problem, by the duality theorem noted earlier. 

On the other hand, conservative auction offers no guarantee of termination: we may end up with a situation where the object prices stop changing, while the cardinality of the current assignment stops growing, as some persons simply change their assigned objects in some way. In particular, by Eqs.\ \bestprofit-\secondbest, the new price $\ol p_{j_i}$ of the preferred object $j_i$ cannot decrease, i.e., 
$$\ol p_{j_i}\ge p_{j_i},$$
 and it will increase strictly (i.e., $\ol p_{j_i}> p_{j_i}$) if and only if the profit $a_{ij_i}-p_{j_i}$ of $j_i$ is strictly larger than the second best profit $w_i$; cf.\ Fig.\ \figconservativepricechange. Thus neither the object prices nor the cardinality of the current assignment will change if there are multiple objects that offer maximum profit for person $i$, and all of these objects are assigned. \old{Thus 
a conservative auction iteration may not change the current prices, and it may also not change the cardinality of the current assignment (this will happen if $j_i$ is already assigned to some person $\ol i$). }

Typically, the cause of nontermination of conservative auction can be traced to what we will call a {\it competitive impasse\/}. We can somewhat loosely describe competitive impasse as a situation where there is a set of persons that compete for a smaller number of (more than one) equally desirable objects, and there is no apparent way to allocate objects to persons without leaving some person(s) dissatisfied in the end; see the example of Fig.\ \figthreebythree. In practice, however, conservative auction can quickly succeed in assigning a substantial number of objects, and for this reason it can be used for effective initialization of other assignment algorithms, as was noted in the papers [Ber81] and [JoV87].

\old{Conservative auction ideas are also central in the relaxation algorithm for the more general minimum cost flow problem and the RELAX code for this problem; see [Ber95], [BeT88], [BeT94]. }

\subsection{Aggressive Auction}
\vskip-0.5pc
\xdef\figaggrpricechange{\figr}\figrnum\show{myfigure}

\pn The aggressive auction algorithm is similar to its conservative counterpart, but guarantees convergence to a complete assignment. In particular, a competitive impasse is resolved by requiring that a bid by an unassigned person $i$ for the best object $j_i$ increases the price of $j_i$ by at least some positive increment $\e$. In particular, person $i$ raises the price of the  best object $j_i$ by the amount
$$\p_i-w_i+\e,$$
where $\p_i=\max_{j\in A(i)}\{a_{ij}-p_j\}$ is the profit of the best object, given by Eq.\ \profit, and $w_i$ is the second best profit, given by Eq.\ \secondbest; see Fig.\ \figaggrpricechange.
We refer to this type of auction as {\it aggressive\/}, because in contrast to the conservative type, it is guaranteed to apply positive price rises (at least $\e$), and it may produce a complete assignment where some of the persons are assigned to a non-maximum profit object. We will also  contrast aggressive auction with the cooperative type of auction algorithm (to be discussed shortly), which aims to first detect a competitive impasse and then resolve it through a process of mutual agreement among the competing persons.

\topinsert
\centerline{\hskip0pc\epsfxsize = 3.8in \epsfbox{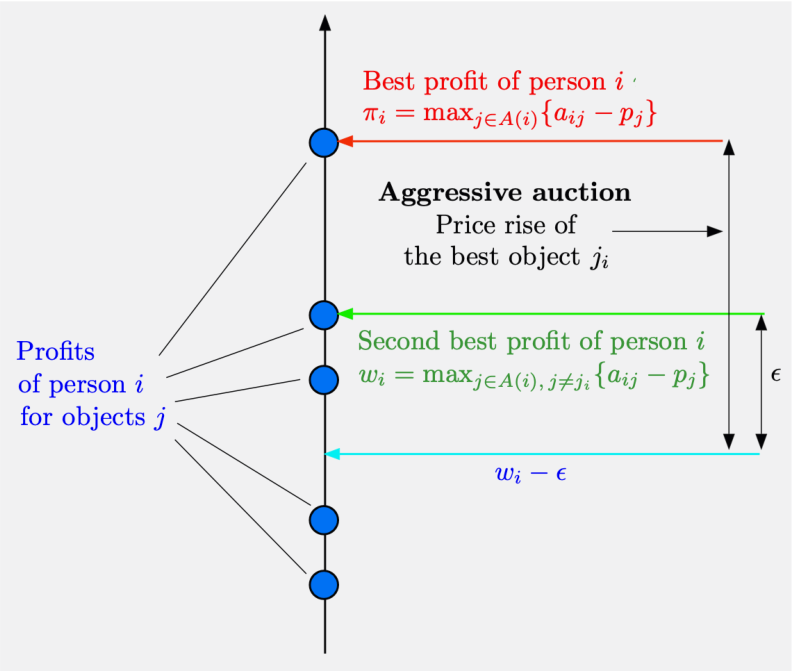}}
\vskip-0.5pc
\hskip-3pc\fig{-0.5pc}{\figaggrpricechange} {Illustration of the price rise of the best object $j_i$ of an unassigned person $i$ in the aggressive auction algorithm. The price of $j_i$ is increased by $\p_i-w_i+\e$, while the profit of $j_i$ is strictly decreased to $w_i-\e$.
}\endinsert

\xdef\figthreebythreeaggr{\figr}\figrnum\show{myfigure}

In summary, given a set of object prices $(p_1,\ldots,p_n)$ and a partial  assignment ${\cal A}$, an aggressive auction iteration generates a new set of prices and a new assignment as described below. The algorithm terminates when following an iteration, the assignment obtained is complete.

\texshopbox{\pn {\bf Aggressive Auction Iteration}
\smskip
\pn We select an unassigned person $i$ and an object $j_i$ that offers maximum profit for $i$ under the given prices,
$$a_{ij_i}-p_{j_i} = \max_{j\in A(i)} \{a_{ij}-p_j\}.\xdef\bestprofit{\lab}\eqnum\show{oneo}$$
 We set the price of $j_i$ to
$$\ol p_{j_i}=a_{ij_i}-w_i+\e\eqnum\show{oneo}$$
where $w_i$ is the ``second best" profit,
$$w_i= \max_{j\in A(i),\, j\ne j_i} \{a_{ij}-p_j\}.$$
Finally, we add to the assignment ${\cal A}$ the pair $(i,j_i)$, and if $j_i$ was assigned to some other person $\tl i$, we remove from ${\cal A}$ the pair $(\tl i,j_i)$, thus forming a new assignment $\bar{\cal A}$.
}

\topinsert
\centerline{\hskip0pc\epsfxsize = 4.0in \epsfbox{three-by-three.eps}}
\vskip-1pc
\vskip0pt
\eightpoint
\def\tablerule{\noalign{\hrule}}
$$\vbox{\offinterlineskip
\hrule
\halign{\vrule\hfill \ #\ \hfill &\vrule\hfill \ #\ \hfill 
&\vrule\hfill \ #\ \hfill &\vrule\hfill \ #\ \hfill 
&\vrule\hfill \ #\ \hfill &\vrule\hfill 
\ #\ \hfill \vrule\cr
&&\cr
{\bf At Start of\lower2ex\hbox{\ }\raise4ex\hbox{\ }}&\hbox{\bf
Object}&\hbox{\bf Assigned}&\hbox{\bf Bidding}&\hbox{\bf Best}&\hbox{\bf Bidding}\cr
{\bf Iteration \#\lower2ex\hbox{\ }}&\hbox{\bf  Prices}&\hbox{\bf  Pairs} &\hbox{\bf Person}&\hbox{\bf Object} &\hbox{\bf Increment}\cr
\tablerule\cr &&\cr
{1\lower2ex\hbox{\ 
}\raise4ex\hbox{\ }\hbox{\ }}&$(0,0,0)$&$(1,1),(2,2)$&3&2&$\e$\cr 
{2\lower2ex\hbox{\
}\raise2ex\hbox{\ }\hbox{\ }}&$(0,\e,0)$&$(1,1),(3,2)$&2&1&$2\e$\cr 
{3\lower2ex\hbox{\
}\raise2ex\hbox{\ }\hbox{\ }}&$(2\e,\e,0)$&$(2,3),(3,1)$&1&2&$2\e$\cr 
{4\lower2ex\hbox{\ 
}\raise2ex\hbox{\ }\hbox{\ }}&$(2\e,3\e,0)$&$(1,2),(2,1)$&3&1&$2\e$\cr 
{5\lower2ex\hbox{\
}\raise2ex\hbox{\ }\hbox{\ }}&$(4\e,3\e,0)$&$(1,3),(3,2)$&2&2&$2\e$\cr 
{6\lower2ex\hbox{\
}\raise2ex\hbox{\ }\hbox{\ }}&$\ldots$&$\ldots$&$\ldots$&$\ldots$&$\ldots$\cr 
&&\cr \tablerule\cr}}$$
\vskip-1pc
\fig{-1pc}{\figthreebythreeaggr} {Illustration of how the aggressive auction algorithm overcomes the competitive impasse problem for the $3\times 3$ example of Fig.\ \figthreebythree\ by making the bidding increment at least equal to $\e$. The table shows one possible sequence
of bids and assignments generated by the auction algorithm, starting with all prices equal to 0 and the partial assignment $\big\{(1,1),(2,2)\big\}$. At each iteration except the last, the unassigned person bids for either object 1 or 2, increasing
its price by $\e$ in the first iteration and by $2\e$ in each subsequent iteration. In the last iteration, after the prices 1 and 2 rise to or above C, object 3 receives a bid and the auction terminates. The number of iterations for this to happen is roughly ${C/\e}$.  
}\endinsert

\figrnum\show{myfigure}

 It can be shown that the algorithm is guaranteed to terminate (under our assumption that there exists at least one complete assignment; see the original paper [Ber79], or the books [BeT89], [Ber91a], [Ber98] for a proof). Intuitively, the reason is that each bid by a person $i$ is guaranteed to increase the price of his/her best object $j_i$ by at least the positive increment $\e$, thus making $j_i$ ``less attractive" for other persons. If the auction did not terminate, the prices of the assigned objects would eventually increase to sufficiently high levels to make some of the unassigned objects attractive enough to receive bids and join the assignment. This is similar to what happens in real-life auctions.
 
The aggressive auction algorithm is designed to maintain the following relaxed form of the CS condition \compslackfinal, called {\it $\e$-complementary slackness}  ($\e$-CS for short):
$$a_{ij_i}-p_{j_i} \ge \max_{j\in A(i)} \{a_{ij}-p_j\}-\e,\qquad \hbox{for all assigned pairs }(i,j_i),\xdef\ecompslacktemp{\lab}\eqnum\show{oneo}$$
provided the initial set of prices and partial assignment satisfy this condition.  One possibility to satisfy the $\e$-CS condition initially is to start with an arbitrary set of prices and the empty assignment. There are also other more sophisticated possibilities for selecting favorable initial conditions.

Thanks to the $\e$-CS condition, it can be shown that the final assignment obtained is optimal within $n\e$, and hence exactly optimal if the values $a_{ij}$ are integers and $\e<1/n$.
To see this, note that the complete assignment and set of prices obtained at termination {\it satisfy CS for a fictitious/slightly perturbed problem} where all values $a_{ij}$ are the same as before, except for the values $a_{ij_i}$ of the $n$ assigned pairs $(i,j_i)$, which are modified by an amount of no more than $\e$; cf.\ Eq.\ \ecompslacktemp. The final complete assignment is optimal for this perturbed problem, and therefore also optimal within $n\e$ for the original (unperturbed) problem.
Thus thanks to the extra $\e$ bidding increment, the  aggressive auction algorithm succeeds in terminating with a complete assignment, at the risk of some persons ending up with a non-maximum profit object (by as much as $\e$), and an attendant error of at most $n\e$ from optimality.

Unfortunately, the aggressive auction algorithm runs into another difficulty, which  can also be traced to a competitive impasse. This difficulty, called a {\it price war\/}, refers to a protracted sequence of small price rises of order $\e$, which results from groups of persons competing for a smaller number of two or more objects that are more or less equally desirable. An example of a price war in the case of a $3\times 3$ assignment problem is given in Fig.\ \figthreebythreeaggr, and it can be seen that it degrades computational efficiency. In particular, the number of iterations in this example is proportional to $C/\e$, 
and a similar example (given as Exercise 7.4b in the book [Ber98]) shows that the number of iterations needed to resolve a price war can be as high as $nC/\e$.

Generally, the complexity of the algorithm can be shown to be proportional to $C/\e$, where 
$$C=\max_{i=1,\ldots,n,\,j\in A(i)}|a_{ij}|\xdef\range{\lab}\eqnum\show{oneo}$$
is the range of object values. Thus, the complexity is pseudopolynomial and is often unacceptable. In actual use of the aggressive auction algorithm, price wars are common, particularly when the range $C$ is large and the assignment problem is sparse, i.e., each person can be assigned to only a small subset of objects).

One way to overcome the detrimental effect of price wars is {\it $\e$-scaling\/}, a natural computational idea that was noted in the original aggressive auction proposal of the paper [Ber79]. Here the algorithm is first run for a fairly large initial value of $\e$, to converge quickly and yield good object price estimates. These estimates are used to initialize an aggressive auction with a reduced value of $\e$. After several successive rounds of $\e$-reduction by some constant factor, this process will bring $\e$ to a sufficiently low level to produce an optimal assignment. It can be shown that the (worst-case) computational  complexity of aggressive auction with $\e$-scaling is polynomial, $O\big(nm\log(nC)\big)$, where $m$ is the number of arcs of the bipartite graph representing the assignment problem and $C$ is the range of values, given by Eq.\ \range. This estimate was derived in the author's  textbook [BeT89] (Section 5.4) and paper [Ber88], following a progression of related complexity analyses for the max-flow and the minimum cost flow problem involving several works (Karzanov [Kar74], Shiloach and Vishkin [ShV82], Goldberg and Tarjan [GoT86], [GoT90], Bertsekas and Eckstein [BeE87], [BeE88], Ahuja, Magnanti, and Orlin [AMO88], [AMO89], Ahuja and Orlin [AhO89], Cheriyan and Maheshvari [ChM89], Orlin and Ahuja [OrA92]).  The recent papers by  Naparstek and Leshem [NaL16], and Khosla and Anand [KhA21] provide probabilistic complexity analyses. 

For an account of the computational complexity aspects of the aggressive auction algorithm with $\e$-scaling, see the textbooks [BeT89] (Section 5.4) and [Ber98] (Section 7.1.2). The latter textbook also contains detailed discussions (including computational complexity) of extensions of the auction algorithm to related problems, such as asymmetric assignment problems, max-flow, minimum cost flow, with both linear (in Chapter 7) and convex separable cost (in Chapter 9).

Aggressive auction with an efficient $\e$-scaling implementation is widely recognized as one of the most effective assignment algorithms.\footnote{\dag}{\ninepoint The experimental verification of the advantages of the aggressive auction algorithm took a long time to establish, owing in part to the primitive state of computer technology at the time. Indeed, given that the aggressive auction algorithm appeared to be radically different  from the established assignment algorithms in 1979, like primal simplex and Hungarian, and lacking a thorough computational comparison, the author harbored deep doubts about its effectiveness. In fact, these doubts prompted the development of an alternative cooperative algorithm (with $\e=0$), which appeared to be conceptually closer to the  Hungarian method, the most popular assignment algorithm at the time; see [Ber81]. The story of the discovery of the aggressive auction algorithm is recounted near the end of a videolecture by the author that can be found at https://www.youtube.com/watch?v=T-fSmSqzcqE} Several  code implementations are publicly available, including some (written in FORTRAN and dating from the early 90s) that can be found in the author's website. A recent code, written in MATLAB, has been made available by Bernard [Ber23a]. The algorithm typically outperforms its competitors by a wide margin, as has been shown convincingly by many computational studies. Its advantage is particularly pronounced when good initial object price estimates are available. As a result, the method is very efficient in situations where many similar assignment problems are solved with small variations in their data. Then the final prices for a given problem solution can be used as starting prices for solution of other similar problems, often with impressive computational savings.\footnote{\dag}{\ninepoint Such situations arise often in practice. An example is data association contexts, where related two-dimensional assignment problems are solved repeatedly; see  the author's monograph [Ber20a] (Section 3.4.2) and  paper [Ber20b], and references on multi-target tracking, such as Blackman [Bla86], Bar-Shalom and Fortman [BaF88], Bar-Shalom [Bar90], Casta\~ non [Cas92], Pattipati, Deb, Bar-Shalom, and Washburn [PDB92], Poore [Poo94], Poore and Robertson [PoR97], Popp, Pattipati, and Bar-Shalom [PPB01], and Emami et al.\ [EPE20].}

Another advantage of the aggressive auction algorithm and its extensions to other network flow problems is that it is well-suited for parallel computation, and it is valid even when it is implemented as a distributed asynchronous algorithm. This has been established in the book by Bertsekas and Tsitsiklis [BeT89] (Sections 5.3 and 6.5), as well as in several related computational studies: Bertsekas and Casta\~ non [BeC91], Wein and Zenios [WeZ91], Amini [Ami94], Bertsekas at al.\ [BCE95], Beraldi, Guerriero, and Musmanno [BGM97], [BGM01], [BeG97], Zavlanos, Spesivtsev, and Pappas [ZSP08], Bus and Tvrdík [BuT09], Sathe, Schenk, and Burkhart  [SSB12],  Nascimento at al.\ [NVJ16], Naparstek  and Leshem [NaL16], Sena, Silva, and Nascimento [SSN21]. The cooperative auction algorithms to be discussed next, can also use good initial price estimates with advantage, but they are not as well suited for distributed computation.

\vskip-1pc

\section{Cooperative Price Rises and Cooperative Auction}
\vskip-0.5pc
\pn We will now consider an alternative approach for dealing with competitive impasses and price wars. The key characteristic that differentiates it from the aggressive auction approach is the use of multiple-object price rises that aim to forestall price wars. In effect, a group of persons recognize that they are caught up in a multi-object competitive impasse, and rather than engage in a time consuming price war, they collectively agree to raise the prices of the relevant objects by a large common increment, thus preparing to bid for additional objects without violating $\e$-CS. 

We call such multi-person bid mechanisms {\it cooperative\/}, and we will show that they can be combined harmoniously with the aggressive and conservative auction mechanisms that involve bids by a single person. In fact, the paper [Ber81] included combinations of cooperative multi-person bids with conservative single-person bids, and experimentally demonstrated the potential advantages of such combinations.

To understand the cooperative auction mechanism, let us consider the $3\times 3$ assignment problem of Figs.\ \figthreebythree\ and \figthreebythreeaggr. There, starting with zero prices, persons  1, 2, and 3 compete for valuable objects 1 and 2 (value $C$), and aim to avoid assignment to the valueless object 3. As we have seen in Fig.\ \figthreebythree, conservative auction fails for this problem, due to a competitive impasse created by perpetual zero-increment  bids by persons 1, 2, and 3, for the two desirable objects 1 and 2. Aggressive auction succeeds in finding the optimal assignment after a protracted price war that lasts for about $C/\e$ iterations, as illustrated in Fig.\ \figthreebythreeaggr. Cooperative auction, aims instead to detect the competitive impasse, to identify the set of persons that are involved in it, and to {\it form a coalition of these persons for the purpose of performing a cooperative price rise} to resolve quickly the  impasse within the coalition. In particular, persons 1, 2, and 3 agree to raise the prices of objects 1 and 2 from 0 to $C+\e$, preserving $\e$-CS, while allowing object 3 to be assigned at the next iteration, thus resolving the competitive impasse without a price war.\footnote{\dag}{\ninepoint In a real auction the person that is ultimately assigned to the valueless object 3 may need to be compensated by prior agreement with his/her coalition partners. This issue is not addressed in this paper, because our objective is computational efficiency in solving the assignment problem, and not the design of fair real-life auction mechanisms. Some possibilities include consideration of profit sharing between persons, or randomized solutions, whereby persons can acquire fractional amounts of multiple objects, with the fractions adding to 1 for each person.} This example also illustrates that {\it price wars involve more than one object\/}. This motivates the use of an aggressive bidding approach when the $\e$-zone of the bidding person contains only one object. We will return to this theme later in this section.

We will now extend the idea just described to the general $n\times n$ assignment problem. To this end we need to address the following issues:

\nitem{(a)} The algorithm should maintain a partial assignment and a set of prices that satisfy CS (or $\e$-CS).  Thus, once the algorithm terminates, the complete assignment obtained at termination is optimal (or optimal within $n\e$, respectively).

\item{(b)} As in the case of conservative and aggressive auctions, the algorithm should aim to enlarge the current partial assignment as long as this is done without violating CS or $\e$-CS.

\nitem{(c)} The algorithm needs an explicit or implicit mechanism to detect that there is a competitive impasse or price war going on.
It also needs a mechanism to identify the coalition of persons that are involved in the price war; this coalition will involve a single unassigned person and $m>1$ assigned persons, and the corresponding assigned $m$ objects for which the $m+1$ persons compete.  

\nitem{(d)} Once a coalition of $m+1$ persons is detected, the prices of the corresponding $m$ assigned objects should be raised simultaneously through a cooperative price rise that does not violate CS or $\e$-CS.   An efficient mechanism to calculate the  cooperative  price rise level should be incorporated into the algorithm.

\smskip

In what follows in this paper, we will aim to design a broad class of algorithms and variations thereof, which are based on the preceding considerations, and mitigate the occurrences of competitive impasses and price wars. To this end we introduce some definitions, all of which refer to a specific set of prices $p=(p_1,\ldots,p_n)$ and partial assignment ${\cal A}$ satisfying $\e$-CS for some fixed $\e\ge0$ (note that $\e=0$ is a possibility). 
If $\e>0$, the algorithm can be combined with $\e$-scaling, i.e., applying the algorithm with larger values of $\e$ to obtain good starting prices for applying the algorithm with smaller values of $\e$. However, the algorithm works even with $\e=0$. 

\subsubsection{Preliminary Concepts}

\pn We first introduce the concept of $\e$-zone of a person, a new idea that plays a central role in this paper.

\xdef\definitionezone{\defn}\defnum\show{myproposition}

\texshopbox{\definition{\definitionezone: ($\e$-Zone of a Person)}Given a set of prices $p$, the {\it maximum profit} of a person $i$, denoted $\p_i$, is defined as
$$\p_i=\max_{j\in A(i)}\{a_{ij}-p_j\}.$$
For a given $\e\ge0$, the {\it $\e$-zone of a person $i$}, denoted ${\cal Z}(i)$, is the set of objects $j$ whose profit for $i$ is within $\e$ of being maximal:
$${\cal Z}(i)=\big\{j\in A(i)\mid a_{ij}-p_j\ge \p_i-\e\big\}.$$
}

\xdef\figezone{\figr}\figrnum\show{myfigure}

\topinsert
\centerline{\hskip0pc\epsfxsize = 3.9in \epsfbox{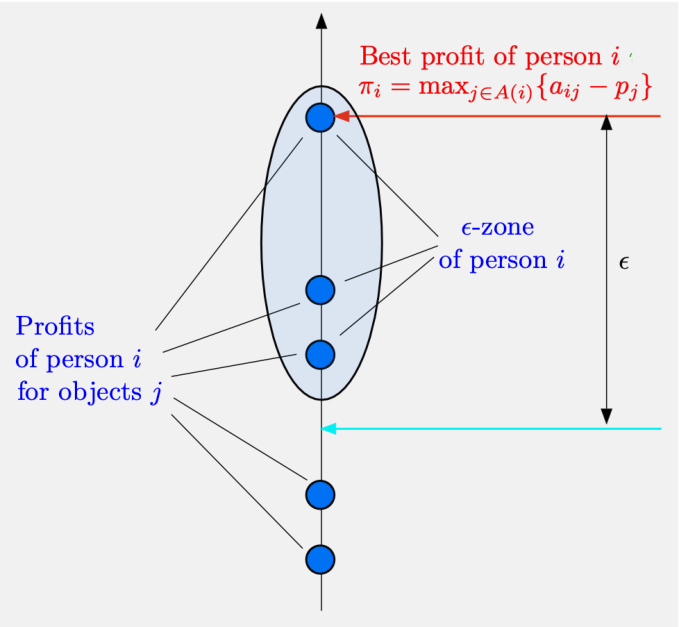}}
\vskip-1pc
\hskip-3pc\fig{-0pc}{\figezone} {Illustration of the $\e$-zone of a person $i$. It consists of all the objects whose profit is within $\e$ of the maximum profit $\p_i=\max_{j\in A(i)}\{a_{ij}-p_j\}$.}\endinsert

Figure \figezone\ illustrates the above definition. Note that the $\e$-zone ${\cal Z}(i)$, roughly speaking, consists of the ``almost best" objects of person $i$ (those whose profit is within $\e$ of being best). It always contains the maximum profit object(s) for person $i$ (and only those if $\e=0$). Moreover, if a person $i$ is assigned to an object $j$ while $\e$-CS holds, then $j$ belongs to the $\e$-zone ${\cal Z}(i)$.

\xdef\definitionaltpath{\defn}\defnum\show{myproposition}

\texshopbox{
\definition{\definitionaltpath\ (Alternating Path):}Let a set of prices $p$ and a partial assignment ${\cal A}$ satisfying $\e$-CS be given. An {\it alternating path} is a person sequence $(i, i_1,\ldots,i_k)$ and corresponding object sequence $(j_1,\ldots,j_k)$, $k\ge 1$,  such that:
\nitem{(a)} The person $i$ is unassigned, while the persons $i_1,\ldots,i_k$ are assigned to objects $j_1,\ldots,j_k$, respectively.
\nitem{(b)} The object $j_1$ belongs to the $\e$-zone of person $i$, while for $m=2,\ldots,k$, the object $j_m$ belongs to the $\e$-zone of person $i_{m-1}$.
}

\xdef\figaltpath{\figr}\figrnum\show{myfigure}

\xdef\definitionaugpath{\defn}\defnum\show{myproposition}

\texshopbox{
\definition{\definitionaugpath\ (Augmenting Path):}Let a set of prices $p$ and a partial assignment ${\cal A}$ satisfying $\e$-CS be given. An {\it augmenting path} is an alternating path $(i, i_1,\ldots,i_k)$, as per Definition \definitionaltpath, together with  an unassigned object $j$ that belongs to the $\e$-zone of $i_{k}$. Given such a path, a corresponding {\it augmentation} consists of assigning person $i$ to $j_1$, reassigning person $i_k$ to $j$,  and reassigning persons $i_1,\ldots,i_{k-1}$ to objects $j_2,\ldots,j_k$, respectively (thereby increasing the cardinality of the assignment by one, while maintaining $\e$-CS). Assigning an unassigned person $i$ to an unassigned object $j$ within his/her $\e$-zone ${\cal Z}(i)$ is also viewed as an augmentation. 
}

\topinsert
\centerline{\hskip0pc\epsfxsize = 2.7in \epsfbox{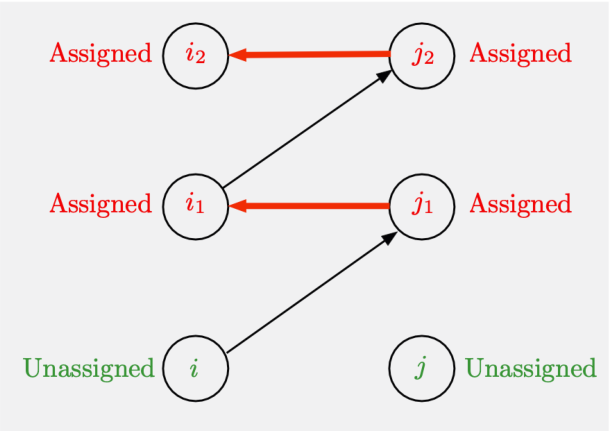}}
\vskip-1pc
\hskip-3pc\fig{-0.5pc}{\figaltpath} {Illustration of an alternating path $(i,i_1,i_2)$, consisting of the three persons in a $3\times 3$ assignment graph. The first person is unassigned and the subsequent persons are assigned. The objects in the alternating path must belong to the $\e$-zones of the corresponding persons in the path, i.e., $j_1\in {\cal Z}(i)$ and $j_2\in Z(i_1)$ [in addition to $j_1\in Z(i_1)$ and $j_2\in Z(i_2)$, which is true by $\e$-CS]. Another alternating path is $(i,i_1)$.}\endinsert

\xdef\figaugpath{\figr}\figrnum\show{myfigure}

An augmenting path as defined above, is denoted by $(i,i_1,\ldots,i_k,j)$, while the corresponding alternating path is denoted by $(i,i_1,\ldots,i_k)$ [in the case where $i$ is assigned to $j$, the augmenting path is denoted $(i,j)$]. Figures \figaltpath\ and \figaugpath\  illustrate alternating and augmenting paths.
Key observations here are that: 
\nitem{(a)} An augmenting path starts with an unassigned person $i$ and ends with an unassigned object $j$, while all other persons and objects in the path are assigned.
\nitem{(b)} Person $i$ and object $j$ can get assigned through an augmentation, which reassigns objects to persons, while maintaining $\e$-CS.
This augmentation makes progress towards obtaining a complete assignment. 
\smskip
\pn The concepts of alternating and augmenting paths are well-known (for the case $\e=0$) in the theory of assignment, matching, and  max-flow algorithms. In particular, an augmentation increases the cardinality of the assignment by 1, while changing the maximal profits of the persons of the augmenting path by no more than $\e$. Thus an augmentation makes intuitive sense for small values of $\e$.

\topinsert
\centerline{\hskip0pc\epsfxsize = 2.7in \epsfbox{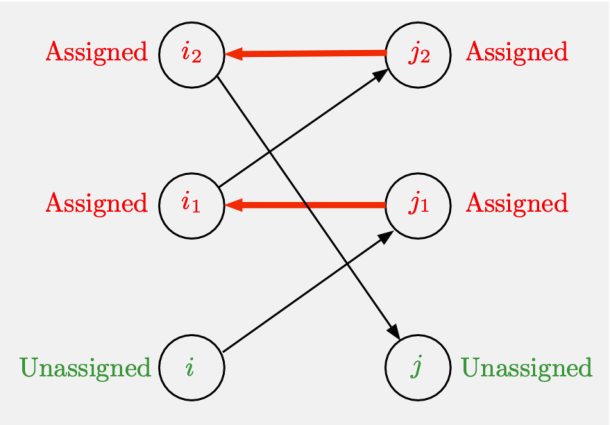}}
\vskip-1pc
\hskip-3pc\fig{-0.5pc}{\figaugpath} {Illustration of an augmenting path  $(i,i_1,i_2,j)$  in a $3\times 3$ assignment graph. It consists of the alternating path $(i,i_1,i_2)$ (cf.\ Fig.\ \figaltpath), followed by the unassigned object $j$, which belongs to the $\e$-zone of person $i_2$.
}\endinsert

\xdef\definitioncoalpart{\defn}\defnum\show{myproposition}

\xdef\figcoalition{\figr}\figrnum\show{myfigure}

We now introduce a notion of coalition of persons, which is central in cooperative auction. 

\vskip-1pc 
\texshopbox{\definition{\definitioncoalpart\ (Coalition Partners of an Unassigned Person):}Let a set of prices $p$ and a partial assignment ${\cal A}$ satisfying $\e$-CS be given, and let $i$ be an unassigned person. A person $i'$ is said to be a {\it coalition partner} of $i$ if there is an alternating path that starts with $i$ and ends with $i'$. The set of persons consisting of  person $i$ together with all his/her coalition partners is called the {\it coalition of $i$} and is denoted by ${\cal C}(i)$. 
}

 Figure \figcoalition\ provides illustrations of ${\cal C}(i)$, the coalition of $i$. Generally, ${\cal C}(i)$ consists of a single unassigned person, namely  $i$, together with $m\ge0$ assigned coalition partners. It consists of the single person $i$ [${\cal C}(i)=\{i\}$] if and only if the $\e$-zone of $i$ does not include any assigned objects (cf.\ the top left graph of Fig.\ \figcoalition). 
We note that we can obtain ${\cal C}(i)$ by using a form of forward search that progressively generates a tree of alternating paths starting from $i$, until no more assigned persons can be found; see the implementation details given later in this section.

\topinsert
\centerline{\hskip0pc\epsfxsize = 4.0in \epsfbox{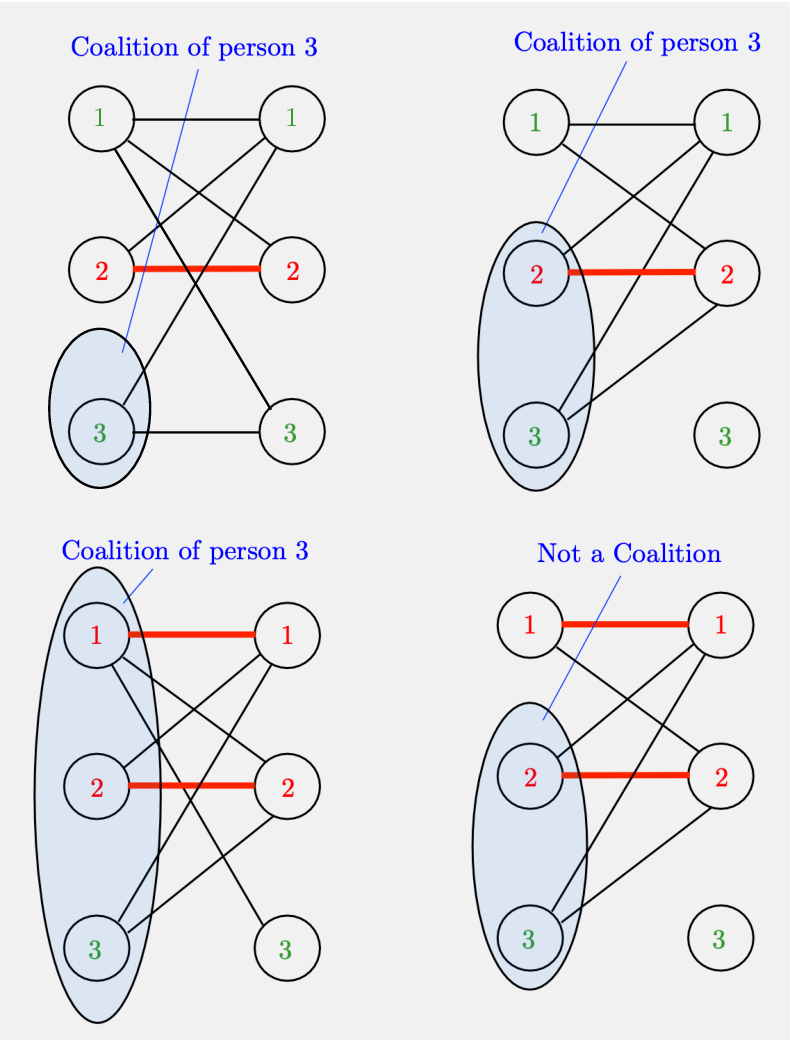}}
\vskip-1pc
\hskip-3pc\fig{-0.5pc}{\figcoalition} {Illustrations of different cases of ${\cal C}(3)$, the coalition of the unassigned person 3 in a $3\times 3$ assignment problem. In each of the four cases, an arc $(i,j)$ indicates membership of object $i$ in the $\e$-zone of person $i$ (other arcs are not shown). Red arcs correspond to assigned pairs, black arcs to unassigned pairs.}\endinsert

\vskip-0.5pc
\subsubsection{Purely Cooperative Auction Iteration}
\vskip-0.5pc
\pn An auction iteration involving a cooperative price rise can now be described in words. We are given a set of object prices $p=(p_1,\ldots,p_n)$ and a partial assignment ${\cal A}$ satisfying $\e$-CS. The iteration starts with an unassigned person $i$ and tries to generate ${\cal C}(i)$, the coalition of $i$.
When ${\cal C}(i)$ is obtained without intermediate discovery of an augmenting path, the prices of all the objects involved in the coalition will be simultaneously raised. Similar to the aggressive auction iteration, the price rise amount exceeds $\e$, and is the maximum possible that preserves $\e$-CS. 

We will now state in detail the iteration just described in summary. We call it ``purely" cooperative, to distinguish it from a method that involves a combination with the conservative and  aggressive iterations. We will describe this combined method later in this section.

\texshopbox{\pn {\bf Purely Cooperative Auction Iteration}
\smskip
\pn Given a set of object prices $p=(p_1,\ldots,p_n)$ and a partial  assignment ${\cal A}$ satisfying $\e$-CS, select an unassigned person $i$.   Let ${\cal M}(i)$ be the set of augmenting paths that start with $i$.
\nitem{$\bullet$} If ${\cal M}(i)$ is nonempty, perform an augmentation along some augmenting path from ${\cal M}(i)$, increase the price of the last object in this augmenting path by the maximum amount that will not violate $\e$-CS, and go to the next iteration.
\nitem{$\bullet$}  If ${\cal M}(i)$ is empty, let ${\cal O}(i)$ denote the set of objects that are assigned to some coalition partner of $i$. Raise the prices of the objects in ${\cal O}(i)$ by the maximum common amount for which the $\e$-zone of every person $i'$ in ${\cal C}(i)$ is a subset of the $\e$-zone of the same person $i'$ after the price rise.}

The preceding iteration description of the cooperative auction algorithm leaves out the details of the computations of the sets ${\cal M}(i)$, ${\cal C}(i)$, and ${\cal O}(i)$, and the price rise amount. To implement efficiently the iteration, it is necessary to properly organize and streamline these computations. The data structures and procedures for doing so are similar to well-known implementations of auction, Hungarian, and dual descent algorithms, and will be discussed later (cf.\ Sections 3.3 and 3.4).
 
\xdef\figcoopauctionexample{\figr}\figrnum\show{myfigure}

\topinsert
\centerline{\hskip0pc\epsfxsize = 3.7in \epsfbox{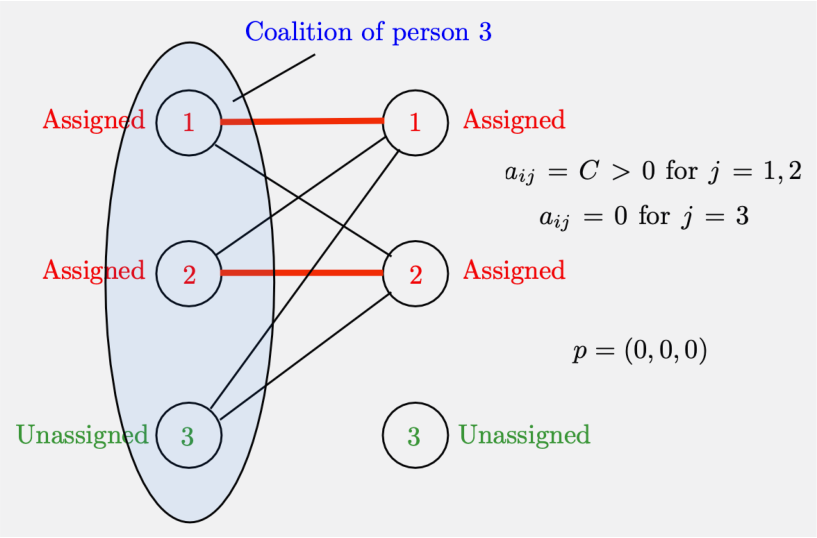}}
\vskip-1pc
\hskip-3pc\fig{-0.5pc}{\figcoopauctionexample} {Illustration of the cooperative auction iteration for the example of Figs.\  \figcoalition, assuming that $C>\e$. Here the coalition partners of the unassigned person 3 are the persons 1 and 2. The cooperative auction iteration consists of a price rise of objects 1 and 2 from 0 to $C+\e$, followed by object 3 coming into the $\e$-zones of all the persons, and allowing an augmentation along $(3,3)$ that completes the assignment.
}\endinsert

Let us illustrate the steps of the cooperative auction iteration with an example.

\xdef\examplethreebythreeone{\exampl}\examplnum\show{myexample}

\beginexample{\examplethreebythreeone}\pn Consider the $3\times 3$ assignment example of  Figs.\ \figthreebythree\ and  \figthreebythreeaggr. We will describe a single iteration of the cooperative auction algorithm, starting with set of prices $p=(0,0,0)$ and partial assignment $\big\{(1,1),(2,2)\big\}$. 

The iteration starts with person 3, the only one left unassigned. We assume that 
$C> \e$, so the $\e$-zone $Z(3)$ is the set of objects $\{1,2\}$. Thus we need to construct the coalition of person 3 with a view towards a cooperative price rise. The alternating paths are $(3,1)$, $(3,2)$, $(3,1,2)$, and $(3,2,1)$, so the coalition partners of person 3 are persons 1 and 2, as illustrated in Fig.\ \figcoopauctionexample. No augmenting path can be found, i.e., ${\cal M}(3)$ is empty, so we increase the prices of objects 1 and 2 by the maximum amount that will not violate $\e$-CS. Thus the prices of objects 1 and 2 are raised to $C+\e$,  adding object 3 to the $\e$-zones of persons 1, 2, and 3. At the next iteration, the augmenting path $(3,3)$ will be discovered, and the algorithm will terminate with an augmentation along $(3,3)$. 

Thus the cooperative auction algorithm terminates very quickly in this example. By contrast, the  conservative auction algorithm would not terminate at all because of a competitive impasse (cf.\ Fig.\ \figthreebythree), while the aggressive auction algorithm would require about $C/\e$ iterations because of a price war (cf.\ Fig.\ \figthreebythreeaggr).\endexample

Note that in the preceding example, the augmenting path $(3,3)$ is created immediately following the price rise, so the corresponding augmentation can be done right away. This suggests a modification of the cooperative auction algorithm so that when an augmenting path is discovered following a price rise, the corresponding augmentation is done right away, rather than wait for another iteration. The expanding coalition variant of the algorithm, which will be discussed shortly, embodies this modification.

If on the other hand an augmenting path is not discovered immediately following a cooperative price rise, there is also a possibility to assign person $i$ through a reassignment of the coalition partners of $i$. We can view this as a somewhat more aggressive form of collective bid of the coalition ${\cal C}(i)$, which aims to  acquire a new object for the coalition, at the expense of deassigning a person from outside the coalition. It leads to another  variant of the cooperative auction algorithm, which will be discussed in Section 4. 

 \vskip-0.5pc
 \subsubsection{Similarities with Noncooperative Auction Iterations}
 \vskip-0.5pc
 
\pn Some similarities with the noncooperative auction iterations, which suggest interesting algorithmic variants, are noteworthy. In particular, assume that the $\e$-zone ${\cal Z}(i)$ contains a {\it single unassigned} object (by necessity the maximum profit object). Then the cooperative auction iteration will produce identical results with the conservative iteration (if $\e=0$) and with the aggressive iteration (if $\e>0$): it will assign $i$ to that object and raise its price by an amount that exceeds $\e$. If on the other hand  the $\e$-zone ${\cal Z}(i)$ contains a {\it single assigned} object $j$, the results will be different, because the person assigned to $j$ is a coalition partner of $i$, and this will trigger the mechanism for computing and enlarging the coalition of $i$. 

In what follows (Section 3.2), we will discuss a variant of the cooperative algorithm that behaves identically with the aggressive auction iteration when  ${\cal Z}(i)$ contains a single object (assigned or unassigned), and is much faster, both in theory and in practice. The motivation for this variant is that the aggressive auction iteration is known to work very fast in the absence of price wars, which involve competition for multiple objects, so {\it a potential price war is not an issue when  ${\cal Z}(i)$ consists of a single object\/}.\footnote{\dag}{\ninepoint For an illustration of why price wars involve at least two objects, consider the $3\times 3$ problem of Fig.\ \figthreebythreeaggr. If there were only one valuable object (value $C$ for all persons) and the other two objects were valueless, the type of price war illustrated in the figure would not occur. For an illustration of how a price war can be generated subsequent to aggressive auction iterations, consider the same example with a fourth person added (with identical values as the other three persons) and a fourth object added offering value $-1$ for all four persons.}  In Section 4 we will describe still another variant of the cooperative algorithm, which behaves identically with the aggressive auction iteration when there is at most one coalition partner of $i$ [rather than ${\cal Z}(i)$ containing a single object]. This is the variant noted earlier, which involves reassignment of the coalition partners of $i$  immediately following a cooperative price rise.
 
 \vskip-0.5pc

 \subsection{Cooperative Auction Iteration With Coalition Expansions}
\vskip-0.5pc
\pn When the augmenting path set ${\cal M}(i)$ is empty (which will happen when all coalition partners of $i$ are assigned), the initial unassigned person $i$ will remain unassigned at the end of the iteration. In this case, since the choice of the unassigned person to start the next iteration is unrestricted, we have the option to start with the same person $i$. Then the new set of coalition partners of $i$ will include the preceding set of coalition partners, so the {\it coalition ${\cal C}(i)$ will be simply expanded and need not be rebuilt from scratch} [by design the $\e$-zone of every person $i'$ in ${\cal C}(i)$ before the price rise is a subset of the $\e$-zone of the same person $i'$ after the price rise].

This observation motivates an interesting variant of the cooperative auction iteration, which involves  multiple successive coalition expansions started by the same single person $i$, up to the point where an augmentation takes place. We call this the {\it expanding coalitions variant\/}, and we note that it will always terminate with an augmentation, resulting in assignment of the starting person $i$, and an increase of the cardinality of the current assignment by 1. Thus, it will produce a complete assignment in exactly $n$ iterations, starting from an empty assignment, while maintaining $\e$-CS throughout the process (under our assumption that the problem is feasible so a complete assignment exists).

\xdef\figfourbyfour{\figr}\figrnum\show{myfigure}

\topinsert
\centerline{\hskip0pc\epsfxsize = 4.5in \epsfbox{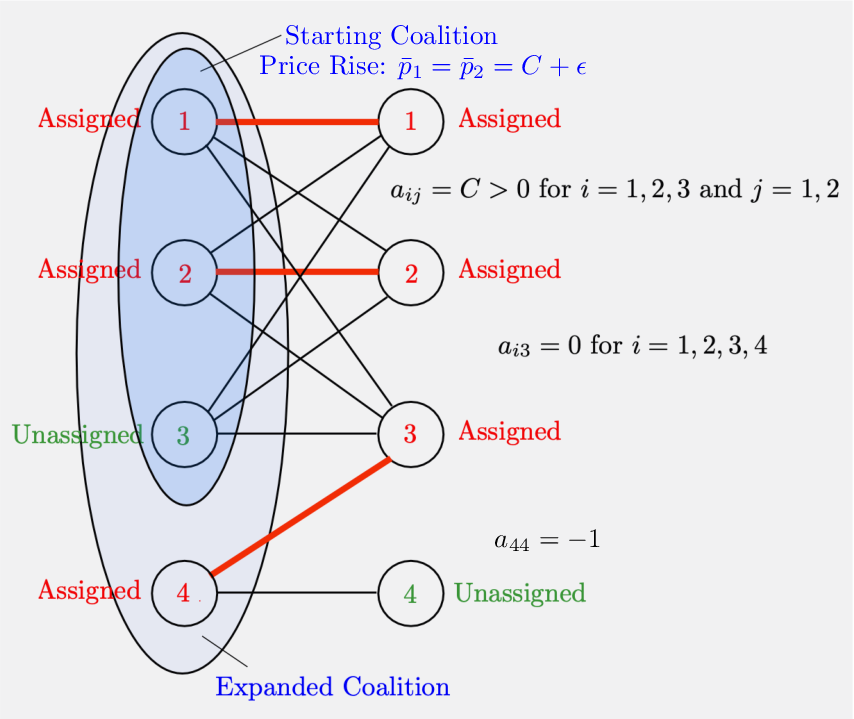}}
\fig{-1pc}{\figfourbyfour} {Illustration of the variant of the cooperative auction iteration that involves an expanding coalition. We consider a $4\times 4$ version of the $3\times 3$ problem of Figs.\ \figthreebythree\ and \figthreebythreeaggr, as shown above. The initial prices are $p=(0,0,0,0)$ and the initial partial assignment is 
$\big\{(1,1),(2,2),(4,3)\big\}.$
We assume that $\e<1/n=1/4$ (to guarantee that the final assignment is optimal). 

The cooperative auction iteration starts with the unassigned person 3, and constructs the coalition of persons 1, 2, and 3, similar to Fig.\ \figcoalition. The prices of objects 1 and 2 rise to $C+\e$, thus bringing the assigned object 3 into the $\e$-zone of the coalition partners 1, 2, and 3. In the expanding coalition variant of the cooperative iteration, the search for coalition partners continues, adding person 4 to the coalition. A new price rise of objects 1, 2, and 3 (by $1+\e$ units) is then performed.  This brings object 4 into the $\e$-zone of person 4 and allows the augmentation $(3,4,4)$, and termination with the  assignment $\big\{(1,1),(2,2),(3,3),(4,4)\big\}$ and prices $\ol p=(C+1+2\e, C+1+2\e,1+\e,0)$.  It can be seen that the final assignment and prices satisfy $\e$-CS.
}\endinsert

\xdef\exampleexpcoalition{\exampl}\examplnum\show{myexample}

\vskip-0.5pc
\beginexample{\exampleexpcoalition}To illustrate the coalition expansion process, let us consider a $4\times 4$ version of the $3\times 3$ problem of Figs.\ \figthreebythree\ and \figthreebythreeaggr. Here, in addition to the three persons and objects of these figures, there are a fourth person 4 and object 4, as shown in Fig.\ \figfourbyfour. Person 4 can be assigned to object 3 with value 0 and to object 4 with value -1. Every feasible assignment must include the pair $(4,4)$, so the optimal assignments are the ones of the $3\times 3$ problem, augmented with $(4,4)$, such as for example
$$\big\{(1,1),(2,2),(3,3),(4,4)\big\}.\xdef\optassign{\lab}\eqnum\show{oneo}$$
Let the initial prices be $p=(0,0,0,0)$ and the initial partial assignment be 
$$\big\{(1,1),(2,2),(4,3)\big\},$$
as shown in Fig.\ \figfourbyfour. The cooperative auction iteration starts with the unassigned person 3, and constructs the coalition ${\cal C}(3)=\{1, 2, 3\}$. In the expanding coalition variant, the search for coalition partners continues after the price rise of objects 1 and 2 (by the amount $C+\e$), adding person 4 to the coalition, which brings object 4 into the $\e$-zone of person 4, and allows the augmentation $(3,4,4)$ and termination with the  assignment \optassign.
\endexample

Here is a more complicated example, which also demonstrates the potentially significant computational savings for reusing the computation of previous coalitions to save in the computation of subsequent coalitions.

\xdef\figexpansion{\figr}\figrnum\show{myfigure}

\xdef\exampleexpansion{\exampl}\examplnum\show{myexample}

\vskip-0.5pc
\beginexample{\exampleexpansion\ (Computational Advantage of Expanding Coalitions)}\pn Consider the $n\times n$ assignment example of  Fig.\ \figexpansion\ (the values $a_{ij}$ are shown above the lines connecting persons and objects). All persons are assigned as shown, except for person $i$, who initiates a cooperative auction iteration with $\e=0$. The starting object prices are $p=(0,\ldots,0)$, and satisfy CS together with the partial assignment shown. 

\topinsert
\centerline{\hskip0pc\epsfxsize = 4.2in \epsfbox{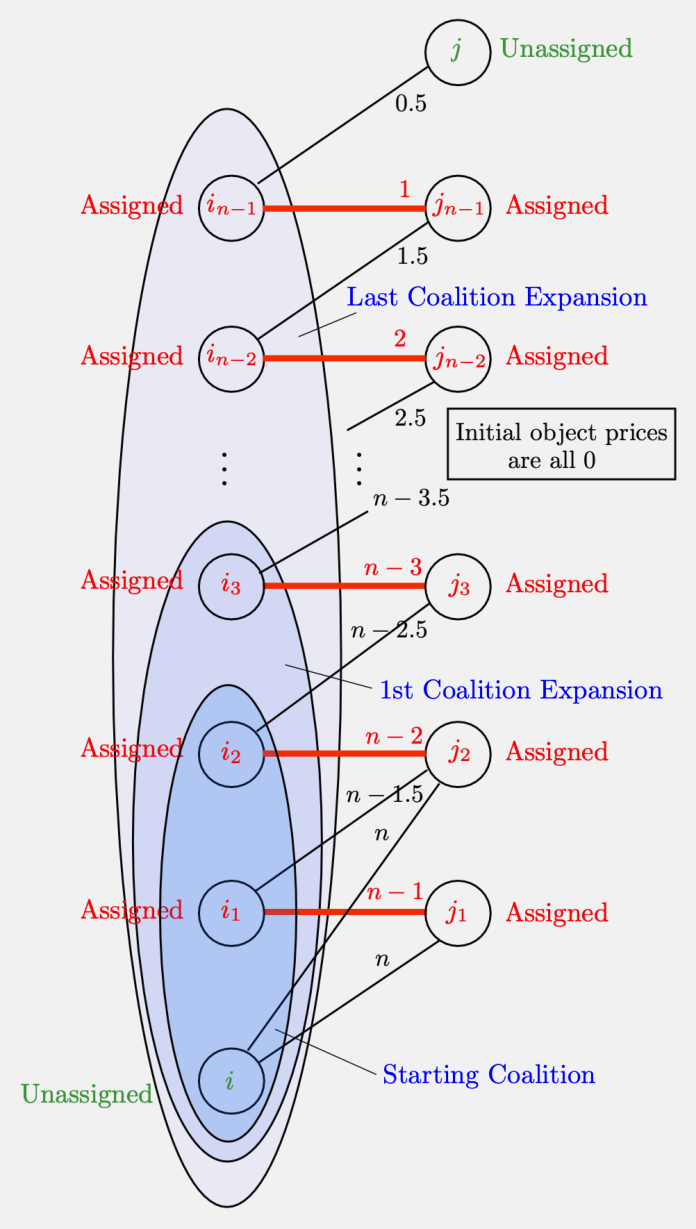}}
\vskip-0pc
\hskip-5pc\fig{-1pc}{\figexpansion} {Illustration of the multiple coalition expansions algorithm with $\e=0$; cf.\ Example \exampleexpansion. The values $a_{ij}$ are shown above the lines connecting persons and objects, and the initial prices are all 0. The algorithm with multiple coalition expansions requires a single iteration (with multiple coalition expansions) and $O(n)$ computation to assign person $i$, while the algorithm without  coalition expansions (cf.\ Section 3) requires $n$ iterations and $O(n^2)$ computation.}\endinsert

Let us apply the cooperative auction algorithm with expanding coalitions and $\e<0.5$. The starting coalition is ${\cal C}(i)=\{i,i_1,i_2\}$ and the price rise of the set of objects ${\cal O}(i)=\{j_1,j_2\}$ is $r=0.5$, which brings object $j_3$ into the 0-zone of person $i_2$. A new iteration is started by person $i$, with coalition ${\cal C}(i)=\{i,i_1,i_2,i_3\}$ and price rise of the set of objects ${\cal O}(i)=\{j_1,j_2,j_3\}$ equal to $r=0.5$, which brings object $j_4$ into the 0-zone of person $i_3$. This coalition expansion process continues for $n-3$ iterations, up to when person $i_{n-1}$ is included in the coalition, the prices of objects $j_1,\ldots,j_{n-1}$ rise by $0.5$, which brings object $j$ into the 0-zone of person $i_{n-1}$, with an augnentation ensuing along the augmenting path $(i,i_2,i_3,\ldots,i_{n-1},j)$. The assignment thus obtained is complete and optimal.

This process requires $n-3$ iterations, and $O(n^2)$ computation (because each of the $n-3$ coalitions is rebuilt from scratch). If it is carried out with the expanding coalition variant, it requires a single iteration with $n-3$ coalition expansions, and $O(n)$ computation.

Suppose now that we use $\e\ge0.5$. Then every object is contained in the $\e$-zone of some person, the starting coalition ${\cal C}(i)$ is the entire person set $\{i,i_1,\ldots,i_{n-1}\}$, and the algorithm terminates in one iteration, without any coalition expansion. There are two possible augmentations from $i$ to $j$: 
$$(i,i_1,i_2, i_3,\ldots,i_{n-1},j)\qquad \hbox{and}\qquad (i,i_2,i_3,\ldots,i_{n-1},j),$$
 and two corresponding complete assignments. The first of these is suboptimal while the second is optimal. The solution generated depends on the order in which persons $i_1$ and $i_2$ enter ${\cal C}(i)$. This illustrates how the number of coalition expansions may be reduced with larger values of $\e$.
\endexample
\vskip-0.5pc

\subsection{Combinations with Noncooperative Auction Algorithms}

\pn We will now explore the possibility of combining the noncooperative auction algorithms (both conservative and aggressive) with the cooperative algorithm. In particular, we are given a set of object prices $p=(p_1,\ldots,p_n)$ and a partial assignment ${\cal A}$ satisfying $\e$-CS. The iteration starts with an unassigned person $i$ and tries to generate the set of coalition partners of $i$. In the process it will perform an aggressive (or conservative) auction iteration if ${\cal Z}(i)$, the $\e$-zone of $i$, contains a single object and $\e>0$ (or $\e=0$, respectively), and a cooperative auction iteration otherwise. The intuitive idea is that {\it when ${\cal Z}(i)$ consists of a single object, there can be a most one coalition partner of $i$, so a price war is not possible\/}. This favors the use of a noncooperative auction iteration.

The iteration just described in summary is stated in detail as follows.

\texshopbox{\pn {\bf Combined Cooperative and Noncooperative Auction Iteration}
\smskip
\pn Given a set of object prices $p=(p_1,\ldots,p_n)$ and a partial  assignment ${\cal A}$ satisfying $\e$-CS, select an unassigned person $i$.   If the $\e$-zone ${\cal Z}(i)$ contains a single object perform a noncooperative auction iteration (conservative if $\e=0$ or aggressive if $\e>0$). Otherwise perform a cooperative auction iteration.}

Note that the iteration can optionally be used with or without coalition expansions. In the former case a cooperative auction iteration is simply continued starting from the same person $i$, up to the point where an augmentation takes place. It should be noted that the purely cooperative auction iteration with coalition expansions and $\e=0$ bears similarity to the Hungarian method, which is typically inferior both in theory and in practice to efficiently implemented aggressive auction iterations. Moreover, its theoretical complexity is known to be inferior to the one of the aggressive auction algorithm. On the other hand, combinations of conservative auction and the Hungarian method have worked well in practice, as verified by the computations given  in the author's paper [Ber81], and by the experience with the JV code [JoV87]. The combined iteration given above, with or without coalition expansions, is new for $\e>0$, and has not been adequately tested, but with proper implementation,  is expected to work more efficiently than either one of its cooperative and noncooperative components working in isolation.

The evaluation of the performance of the combined aggressive and cooperative auction iteration, with $\e>0$ and the expanding coalition process, in conjunction with $\e$-scaling, is an issue of great interest, both theoretically and experimentally. A relevant fact here is that two-phase algorithms, which involve aggressive auction ($\e>0$) in the first phase and  a Hungarian-like  algorithm ($\e=0$) in the second phase after most of the objects have been assigned, have been shown to have computational complexity that is superior to either aggressive auction or the Hungarian method in isolation of each other. In particular, Orlin and Ahuja [OrA92] have derived a related 
$$O\big(\sqrt{n}\,m\log(nC)\big)\xdef\complexity{\lab}\eqnum\show{oneo}$$
 worst-case complexity result for a two-phase algorithm of this type, with the threshold for switching between the two phases skillfully chosen (see also Chapter 5, Exercise 4.5 of the book [BeT89], with solution included in the internet-posted version of the book). The use of $\e>0$ together with $\e$-scaling, requires fewer coalition expansions, as can be seen from Example \exampleexpansion, and seems to be a natural alternative way to deal with a large number of coalition expansions for many problems. Thus it is reasonable to conjecture that a complexity estimate like the one of Eq.\ \complexity\ can be proved for some version of the combined aggressive and cooperative auction iteration.

\subsection{Properties of the Cooperative Auction Algorithm}

\pn In this section we will discuss some general properties and implementations of cooperative auction.
We first note that if there is no augmenting path starting from $i$ [i.e., ${\cal M}(i)$ is empty], the set of objects that are assigned to some coalition partner of $i$, is the union of the $\e$-zones of the persons in the coalition of $i$:
$${\cal O}(i)=\cup_{i'\in{\cal C}(i)}{\cal Z}(i').\eqnum\show{oneo}$$
To see this, note that  when ${\cal M}(i)$ is empty, all objects in the $\e$-zones of $i$ and his/her coalition partners must be assigned to some coalition partner of $i$ (otherwise an augmentation would be performed).

\xdef\figpriceincr{\figr}\figrnum\show{myfigure}
\xdef\figcoopvariant{\figr}\figrnum\show{myfigure}

Let us now provide an explicit formula for the common price rise for the case where ${\cal M}(i)$ is empty. For each person $i'\in{\cal C}(i)$, consider the scalar
$$r_{i'}=\e+\min_{j\in {\cal Z}(i')}\{a_{i'j}-p_j\}-\max_{j\notin {\cal O}(i),\, j\in A(i')}\{a_{i'j}-p_j\},\xdef\pricerise{\lab}\eqnum\show{oneo}$$
(by convention, the maximum above is $-\infty$ if the set over which the maximum is taken is empty). It can be seen from Fig.\ \figpriceincr\ that $r_{i'}$ is the maximum price rise of the objects in ${\cal O}(i)$ that will keep every object in ${\cal Z}(i')$, the $\e$-zone of $i'$, within ${\cal Z}(i')$ following the price rise. To keep {\it all} the objects in ${\cal O}(i)=\cup_{i'\in{\cal C}(i)}{\cal Z}(i')$ within the $\e$-zone of either $i$ or some coalition partner of $i$, the common price  rise should not exceed any one of the amounts $r_{i'}$, $i'\in{\cal C}(i)$. Thus the maximum possible common price rise is  
$$r=\min_{i'\in{\cal C}(i)}r_{i'}.\xdef\minpricerise{\lab}\eqnum\show{oneo}$$

\topinsert
\centerline{\hskip0pc\epsfxsize = 5.4in \epsfbox{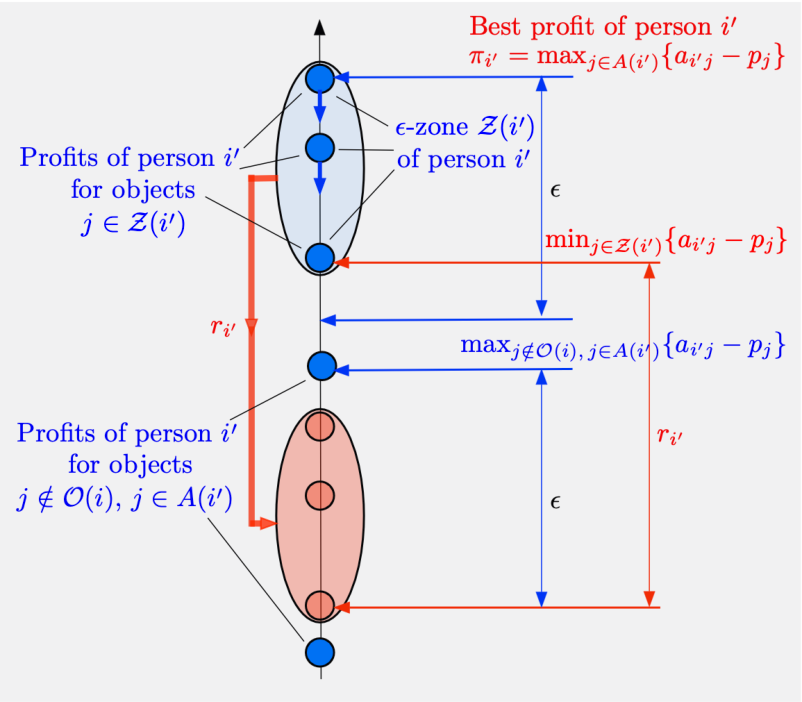}}
\vskip-0.5pc
\hskip-2pc\fig{-0.5pc}{\figpriceincr} {Illustration of a price rise (and corresponding profit drop) of the objects in the $\e$-zone ${\cal Z}(i')$ of a person $i'\in{\cal C}(i)$. The figure shows the maximum amount $r_{i'}$ by which we can raise the prices of objects $j$ in the $\e$-zone ${\cal Z}(i')$, while guaranteeing that all $j\in {\cal Z}(i')$ will stay within the $\e$-zone of $i'$ following a cooperative price rise. It is given by 
$$r_{i'}=\e+\min_{j\in {\cal Z}(i')}\{a_{i'j}-p_j\}-\max_{j\notin {\cal O}(i),\, j\in A(i')}\{a_{i'j}-p_j\},$$
cf.\ Eq.\ \pricerise. 

In the above figure the $\e$-zone ${\cal Z}(i')$ consists of the three most profitable objects of person $i'$ with profits within the top ellipse. The figure also shows the profits of the objects $j\notin {\cal O}(i)$ with $j\in A(i')$ [the profits of any additional objects $j\in {\cal O}(i)$ with $j\in A(i')$ but $j\notin {\cal Z}(i')$ are not shown]. The figure assumes that the set 
$$\big\{j\mid j\notin {\cal O}(i),\, j\in A(i')\big\}$$
 is nonempty; if it is not, we have $r_{i'}=\infty$. 

After we raise the  prices of the objects in ${\cal O}(i)$ by $r_{i'}$, the profits of objects in ${\cal Z}(i')$ move downward by $r_{i'}$, just within $\e$ of the fourth object, which now becomes the most profitable. Price rises by amounts  smaller than $r_{i'}$ still keep the three most profitable objects within the $\e$-zone ${\cal Z}(i')$, but may not be sufficient to bring the fourth object into ${\cal Z}(i')$.}\endinsert

Moreover, following the price rise, the union of the $\e$-zones of the persons in ${\cal C}(i)$  consists of ${\cal O}(i)$ and a nonempty set $\ol {\cal O}(i)$ of additional objects. This is the set of objects that attain the maximum in the maximization [cf.\  Eq.\ \pricerise]
$$\max_{j\notin {\cal O}(i),\, j\in A(i^{''})}\{a_{i^{''}j}-p_j\},$$
while $i^{''}$ attains the minimum in Eq.\ \minpricerise. This shows that the set of objects $\ol {\cal O}(i)$ obtained at the end of the iteration is nonempty and that $r$ is finite (otherwise the existence of a complete assignment assumption would be violated).
Note that if any of the objects within $\ol {\cal O}(i)$, say object $j$, is unassigned, we can perform an augmentation that starts at $i$ and ends at $j$, and increase the price of $j$ by the maximum amount that will not violate $\e$-CS. This can be done efficiently, and it is generally recommended, as it increases the cardinality of the assignment by one, but for simplicity, we have not stated this explicitly. Alternatively, we may suitably modify the cooperative iteration description, so an augmentation is automatically performed, if possible, following a price rise.
Figure \figcoopvariant\ provides an illustration.

\topinsert
\centerline{\hskip0pc\epsfxsize = 3.6in \epsfbox{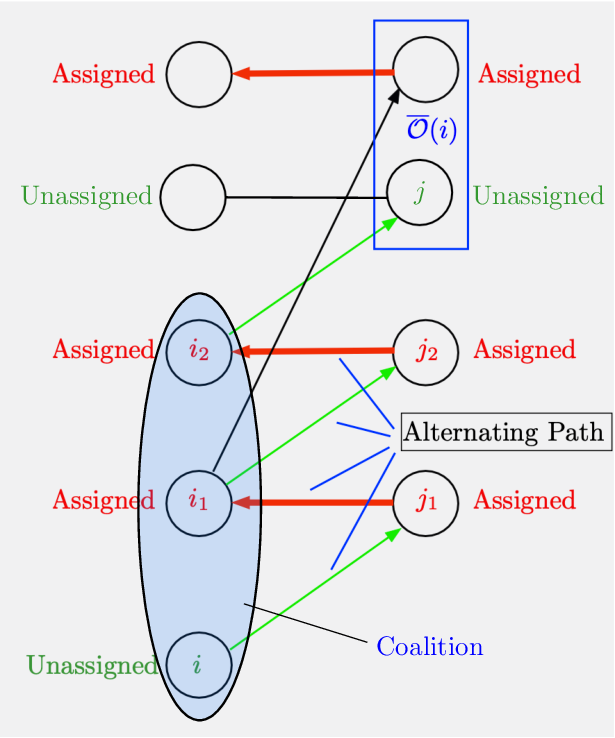}}
\vskip-1pc
\hskip-3pc\fig{-0.5pc}{\figcoopvariant} {Illustration of the cooperative auction iteration, where after raising the prices, there is an object in the set $\ol {\cal O}(i)$ that is  unassigned. Then, we may optionally perform an augmentation, shown in green, along a corresponding augmenting path [$(i,i_1,i_2,j)$ in the figure]. In particular, persons $i$, $i_1$, and $i_2$ get assigned to $j_1$, $j_2$, and $j$, respectively.}
\endinsert

We summarize the principal conclusions from the preceding discussion in the following proposition.

\xdef\propproperties{\propn}\propnum\show{myproposition}

\texshopboxnb{\proposition{\propproperties:} Consider the cooperative auction iteration under the assumption that  there is no augmenting path starting from $i$, i.e., ${\cal M}(i)$ is empty. Consider also ${\cal O}(i)$, the set of objects that are assigned to some coalition partner of $i$. Then following a price rise:
\nitem{(a)} ${\cal O}(i)$ is equal to the union of the $\e$-zones of all persons in ${\cal C}(i)$.}\texshopboxnt{
\nitem{(b)} The prices  of the objects in ${\cal O}(i)$ are raised by the common increment
$$r=\min_{i'\in{\cal C}(i)}r_{i'},$$
where $r_{i'}$ is given by Eq.\ \pricerise, and we have $r>\e$.
\nitem{(c)} Following the price rise, the union of the $\e$-zones of the persons in ${\cal C}(i)$  consists of ${\cal O}(i)$ and a nonempty set $\ol {\cal O}(i)$ of additional objects.
\nitem{(d)} If any of the objects within $\ol {\cal O}(i)$, say object $j$, is unassigned, an augmentation that starts at $i$ and ends at $j$ can be performed. Moreover, the prices and assignment obtained following this augmentation satisfy $\e$-CS.
}

The computational complexity of the algorithm is not expected to be better than the one of the aggressive auction algorithm [$O\big(nm\log(nC)\big)$, where $m$ is the number of arcs of the bipartite graph representing the assignment problem and $C$ is the range of values, given by Eq.\ \range]. However, depending on the implementation and the type of problem addressed, it appears that the cooperative auction algorithm, as given in this section, can outperform the aggressive auction algorithm, particularly in situations where price wars are likely.

\subsection{Common Price Increment Computation}

\pn We will now focus on the most complicated part of a cooperative auction iteration, namely the computation of the common price rise increment $r$ of Eqs.\ \pricerise-\minpricerise, and the new set of objects $\ol {\cal O}(i)$ that are subsequently brought into the coalition of $i$, when there is no augmenting path starting from $i$. We will first describe one possible implementation that can be interpreted graphically, and we will subsequently provide a more general implementation in pseudocode. We assume that we are given a set of object prices $p=(p_1,\ldots,p_n)$ and a partial  assignment ${\cal A}$ satisfying $\e$-CS, together with an unassigned person $i$ to start the iteration. We also assume that the $\e$-zone ${\cal Z}(i)$ contains multiple objects all of which are assigned, and that no augmenting path starting from $i$ exists. 

\xdef\figsearchtree{\figr}\figrnum\show{myfigure}

In particular, we will use the layered graph shown in Fig.\ \figsearchtree\ to illustrate the computation of:

\nitem{(a)} The set of coalition persons ${\cal C}(i)$.

\nitem{(b)} The set of coalition objects ${\cal O}(i)$, i.e., the set of objects assigned to the persons in ${\cal C}(i)$.

\nitem{(c)} The common price rise $r$ of the objects in ${\cal O}(i)$.
\smskip

\pn In this computation we break down the sets ${\cal O}(i)$ and ${\cal C}(i)$ into layers of disjoint subsets
${\cal O}_1,{\cal C}_1,\ldots,{\cal O}_k,{\cal C}_k$,
where for   some positive integer $k<n$, and for $m=1,\ldots,k$:
\nitem{}The $m$th person layer ${\cal C}_m$ is the set of persons $i'$ such that every alternating path that starts at $i$ and ends  at $i'$ contains at least $m$ persons other than $i$.
\nitem{}The $m$th object layer ${\cal O}_m$ is the set of objects that are assigned to the persons in ${\cal C}_m$.
\smskip
\pn The layers ${\cal O}_m$ and ${\cal C}_m$ are computed successively, and can be visualized in terms of the tree of alternating paths shown in Fig.\ \figsearchtree. The details of the computation are as follows:

\texshopbox{\pn{\bf Layer Construction}
\nitem{(a)} We  construct ${\cal O}_1$, which is the set of objects in the $\e$-zone ${\cal Z}(i)$ of person $i$, and then ${\cal C}_1$, which is the set of persons assigned to the objects in ${\cal O}_1$.
\nitem{(b)} Given ${\cal C}_m$, we construct ${\cal O}_{m+1}$ as the set of objects $j\notin {\cal O}_{1}\cup\cdots\cup {\cal O}_{m}$ that belong to the $\e$-zone of at least one person in ${\cal C}_m$; if ${\cal O}_{m+1}$ is empty, then we stop (i.e., $m=k$), having computed ${\cal O}(i)$ and ${\cal C}(i)$ according to
$${\cal O}(i)={\cal O}_1\cup\cdots\cup{\cal O}_k,\qquad  {\cal C}(i)=\{i\}\cup {\cal C}_1\cup\cdots\cup{\cal C}_k.$$}

\topinsert
\centerline{\hskip0pc\epsfxsize = 5.6in \epsfbox{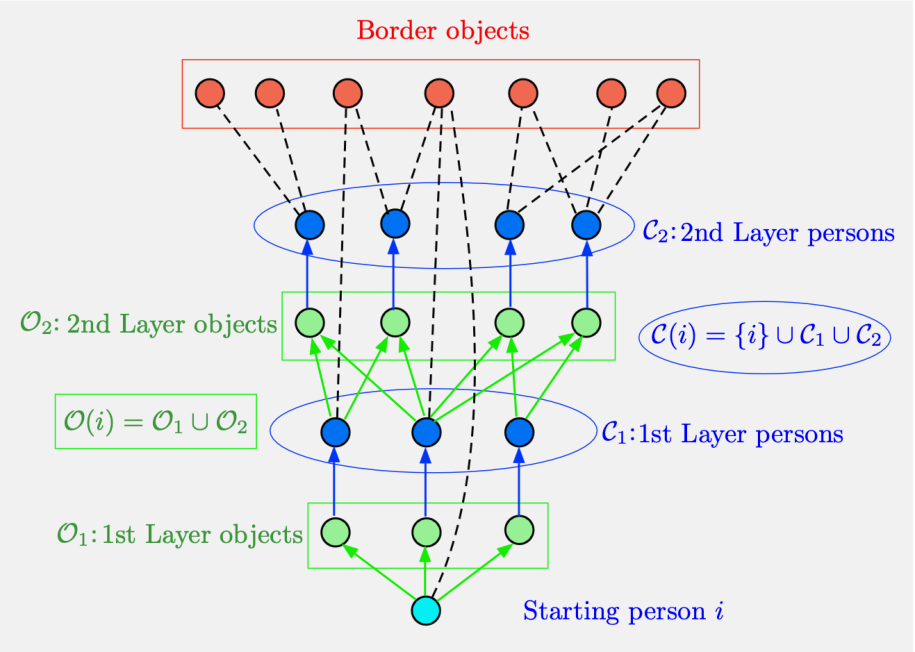}}
\vskip-1pc
\hskip-2pc\fig{-0.5pc}{\figsearchtree} {Illustration of a search tree to construct the coalition of $i$, 
$${\cal C}(i)=\{i\}\cup{\cal C}_1\cup {\cal C}_2,$$
assuming no augmentation occurs during the cooperative auction iteration (the figure assumes two object layers ${\cal O}_1, {\cal O}_2$,  and two person layers ${\cal C}_1, {\cal C}_2$). The objects in ${\cal O}_1$ are the ones in the $\e$-zone ${\cal Z}(i)$. The objects in ${\cal O}_2$ are the ones that do not belong to ${\cal O}_1$ but belong to the $\e$-zone ${\cal Z}(i')$ of some person of ${\cal C}_1$. The set ${\cal B}$ of border objects consists of all objects $j$ that do not belong to 
 $${\cal O}(i)={\cal O}_1\cup {\cal O}_2,$$
 but can be matched with some person $i'\in{\cal C}(i)$, i.e.,
$${\cal B}=\big\{j\notin {\cal O}(i)\mid j\in A(i') \hbox{ for some }i'\in{\cal C}(i)\big\}.$$
Green arrows indicate pairs $(i',j')$ such that $i'\in {\cal C}(i)$ and $j'\in {\cal Z}(i')$. Broken lines indicate pairs $(i',j')$ such that $i'\in {\cal C}(i)$, $j'\in A(i')$ but $j'\notin {\cal O}(i)$. The paths from $i$ to the (blue) nodes in ${\cal C}_1\cup {\cal C}_2$ are the shortest alternating paths.}\endinsert 

In the process of constructing the layers ${\cal O}_1,{\cal C}_1,\ldots,{\cal O}_k,{\cal C}_k,$ we obtain the set of {\it border objects\/}, denoted ${\cal B}$, and consisting of the objects that do not belong to ${\cal O}(i)$ but can be matched with a person in the coalition $ {\cal C}(i)$, i.e.,
$${\cal B}=\big\{j\notin {\cal O}(i) \mid j\in A(i')\hbox{ for some }i'\in {\cal C}(i)\big\};$$
see Fig.\ \figsearchtree. The border objects are obtained during the process of constructing the sets ${\cal O}(i)$ and ${\cal C}(i)$ as described earlier. 

Simultaneously with the computation of ${\cal O}(i)$, ${\cal C}(i)$, and ${\cal B}$ as described above, we can also compute the cooperative price rise amount of the iteration using Eq.\ \minpricerise: 
$$r=\e+\min_{i'\in{\cal C}(i)}\left\{\hat \p_{i'}+\min_{j\in {\cal B},\, j\in A(i')}\{p_j-a_{i'j}\}\right\},\xdef\twomin{\lab}\eqnum\show{oneo}$$
where $\hat \p_{i'}$ is given by
$$\hat \p_{i'}=\min_{j\in {\cal Z}(i')}\{a_{i'j}-p_j\}.$$
Combining the preceding equations with Eq.\ \minpricerise\ and interchanging the order of minimizations in Eq.\ \twomin, we obtain
$$r=\e+\min_{j\in {\cal B}}\min_{i'\in {\cal C}(i),\, j\in A(i')}\big\{\hat \p_{i'}+p_{j}-a_{i'j}\big\}=\e+\min_{j\in{\cal B}}d_{j},\xdef\scalarr{\lab}\eqnum\show{oneo}$$
where for all $j\in{\cal B}\,$
$$d_{j}=\cases{\min_{i'\in {\cal C}(i)}\{\hat \p_{i'}+p_{j}-a_{i'j}\}&if  $j\in A(i')$ for some $i'\in {\cal C}(i)$,\cr
\infty&otherwise.\cr}$$
To understand the intuitive meaning of $d_{j}$, we first note that $\hat \p_{i'}$ is the profit of  person $i'$, assuming $i'$ is awarded the least profitable of the objects in his/her $\e$-zone. Then we can view $d_{j}$ as a {\it profit loss} incurred when person $i'$ is reassigned to $j$ from his/her least profitable object within ${\cal Z}(i')$. The common price rise $r$ of Eq.\ \twomin\ can be interpreted as $\e$ plus the minimum possible profit loss some person $i'$ is reassigned to some $j\in{\cal B}$ from his/her least profitable object in ${\cal Z}(i')$. Note also that each reassignment of a person $i'\in {\cal C}(i)$ to an object  in ${\cal O}(i)$, in the course of an augmentation, involves a loss or gain in profit of at most $\e$, since the objects assigned to $i'$ before and after the augmentation both belong to the $\e$-zone ${\cal Z}(i')$.

Note that $\hat \p_{i'}$  can be computed while we go over the set of associated objects $A(i')$ of person $i'$, to determine whether they can be added to ${\cal O}(i)$. 
Thus the computation of $r$ can be organized progressively: first update the quantity $d_{j}$, as new persons $i'$ are added to the coalition $ {\cal C}(i)$, and then at the end of the iteration, after $ {\cal C}(i)$ and ${\cal B}$ are obtained, take the minimum  over $j\in{\cal B}$ of  $d_{j}$ to obtain $r$; cf.\ Eq.\ \scalarr. Also the set of objects $\ol{\cal O}(i)$ that enter the $\e$-zone of at least one person in the coalition ${\cal C}(i)$ following the price rise, include the ones that attain the minimum of $d_{j'}$ over $j'\in{\cal B}$.

\subsubsection{A More General Implementation of the Coalition Construction Process}

\pn Let us now provide pseudocode for a more general implementation of the cooperative auction iteration that constructs the sets ${\cal C}(i)$, ${\cal B}$, and the scalar $r$ of Eq.\ \scalarr. The code uses two temporary lists of persons $C$ and $C'$. At the end of the iteration, $C={\cal C}(i)$ and $C'=\emptyset$.

\texshopbox{\pn {\bf Pseudocode to Construct the Sets ${\cal C}(i)$ and ${\cal B}$}
\smskip
\pn Initialization: $C=\emptyset$, $C'=\{i\}$, ${\cal B}=\{1,\ldots,n\}$, $d_{j}=\infty$ for all $j\in\{1,\ldots,n\}$.
\pn Until $C'=\emptyset$:
\item{}Remove a person $i'$ from $C'$ and add it to $C$. Let $\hat \p_{i'}=\min_{j\in {\cal Z}(i')}\{a_{i'j}-p_j\}.$ For all $j\in A(i')\cap {\cal B}$:
\itemitem{$\bullet$}If $j\in {\cal Z}(i')$ and $j$ is unassigned, perform an augmentation that starts at $i$ and ends at $j$ and go to the next iteration; otherwise, if $j\in {\cal Z}(i')$ and $j$ is assigned to a person $i''$, remove $j$ from ${\cal B}$, and add $i''$ to $C'$ if it is not already in $C'$.
\itemitem{$\bullet$}If $j\notin {\cal Z}(i')$, set $d_{j}\leftarrow \min\{d_{j},\,\hat \p_{i'}+p_{j}-a_{i'j}\}$. 
\pn Set ${\cal B}\leftarrow \{j\in {\cal B}\mid d_{j}<\infty\}$, ${\cal C}(i)=C$, $r=\e+\min_{j\in{\cal B}}d_{j}.$
}

It can be verified that different rules for choosing the person $i'$ to be removed from $C'$ will lead to the same sets ${\cal C}(i)$ and ${\cal B}$, and the same price rise $r$. On the other hand, one may or may not obtain the layered structure illustrated in Fig.\ \figsearchtree, which corresponds to a special rule for choosing $i'$. This is the rule that removes the persons $i'$ from $C'$ in the same order in which they entered $C'$. Other rules may also be considered based on a heuristic or more principled rationale in a given problem.

\vskip-1pc
\section{Additional Cooperative Auction Variants}
\vskip-0.5pc

\pn 
There are a number of interesting variations of the cooperative auction algorithm, in addition to those we have discussed so far. Most of these variations  are aimed at accelerating convergence, mitigating as much as possible the effects of price wars, and enhancing the suitability for parallel computation. 
Several of these variations have similar theoretical properties. However, their practical performance may be significantly affected by the character of the specific problem that is being solved, such as graph density/sparsity, large/small range of values $a_{ij}$, and special characteristics of the graph's structure, such as large/small ``diameter" (a measure of the average number of hops between two randomly chosen persons).

The wide choice between conservative, aggressive, and cooperative algorithms, and variations thereof, suggests a view of an auction algorithmic landscape where there is no universal best choice that works optimally for all problems. Instead the appropriate choice depends on the characteristics of the type of problem at hand. This view is supported by extensive computational results in the paper [Ber95b], which tested comparatively some (but by no means all) of the algorithmic ideas discussed in the present paper within a broader context of network optimization problems.

 \vskip-0.5pc
 
 \subsubsection{Cooperative Auction Iteration With Collective Bidding and Person Reassignments}

\xdef\figrearrangementsvariant{\figr}\figrnum\show{myfigure}

\pn This variant of the cooperative auction iteration aims to bring it closer to the aggressive auction iteration, at the expense of foregoing the option of expanding coalitions. Consider the cooperative iteration for the case where there is no  augmenting path [${\cal M}(i)$ is empty]. Then after the subsequent collective price rise, the union of the $\e$-zones of the persons in ${\cal C}(i)$  consists of ${\cal O}(i)$ and a nonempty set $\ol {\cal O}(i)$ of additional objects, as we have discussed in Section 3.3. There are now two possibilities:

\nitem{(a)} {\it There is an unassigned object $\bar j$ within the set $\ol {\cal O}(i)$\/}. Then as we discussed earlier, an augmenting path is created following the price rise,  which starts at $i$ and ends at $\bar j$ (cf.\ Example \examplethreebythreeone\  and Fig.\ \figcoopvariant). This augmentation can be performed immediately, without waiting for the next iteration to discover it. 

\nitem{(b)} {\it All objects in the set $\ol {\cal O}(i)$ are assigned\/}. In this case, the purely cooperative auction algorithm of Section 3 simply goes to the next iteration. However, there is also a possibility to assign person $i$ through a reassignment of the coalition persons and a rearrangement of the corresponding assigned pairs. This is illustrated in Fig.\ \figrearrangementsvariant, which should be contrasted with  Fig.\ \figcoopvariant.
\smskip

\topinsert
\centerline{\hskip0pc\epsfxsize = 3.1in \epsfbox{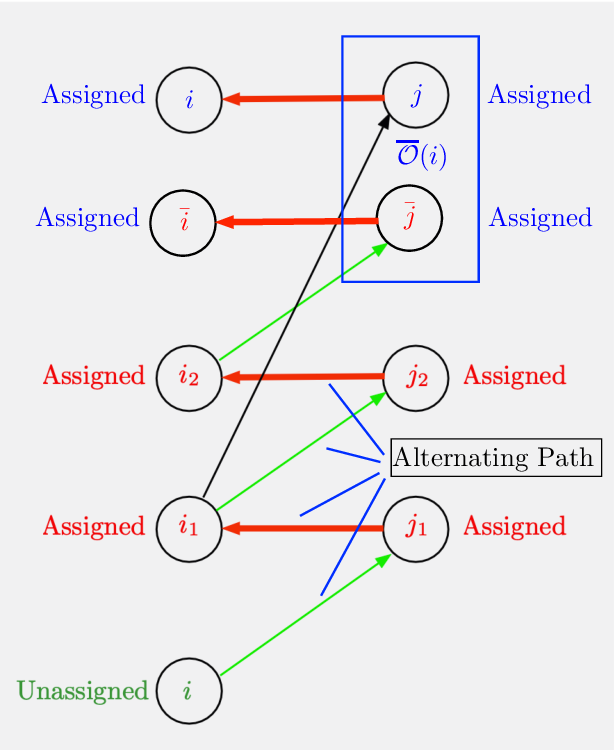}}
\vskip-1pc
\hskip-3pc\fig{-0.5pc}{\figrearrangementsvariant} {Illustration of the cooperative auction iteration with person reassignments when the set of augmenting paths ${\cal M}(i)$ is empty, and all the objects in the set $\ol {\cal O}(i)$ are assigned. Then, we choose an object $\bar j\in\ol {\cal O}(i)$ and perform a reassignment of persons to objects, shown in green, along a corresponding alternating path [$(i,i_1,i_2)$ in the figure]. In particular, persons $i$, $i_1$, and $i_2$ get assigned to $j_1$, $j_2$, and $\bar j$, respectively, while the person $\bar i$ that is assigned to $\bar j$ under ${\cal A}$ becomes unassigned. Note that the object $\bar j$ is not unique: any object in $\ol {\cal O}(i)$ (such as $j$ in the figure) and alternating path corresponding to that object [such as $(i,i_1)$ in the figure] can be used.}
\endinsert

The corresponding auction algorithm variant is identical to the cooperative auction iteration of Section 3, except for the additional person reassignment process, which involves $\ol {\cal O}(i)$ and is performed at the end when ${\cal M}(i)$ is empty. We state this variant formally as follows:

\texshopboxnb{\pn {\bf Cooperative Auction Iteration With Collective Bidding and Person Reassignments}
\smskip
\pn Given a set of object prices $p=(p_1,\ldots,p_n)$ and a partial  assignment ${\cal A}$ satisfying $\e$-CS, select an unassigned person $i$.   Let ${\cal M}(i)$ be the set of augmenting paths that start with $i$.
\nitem{$\bullet$} If ${\cal M}(i)$ is nonempty, perform an augmentation along some augmenting path from ${\cal M}(i)$, increase the price of the last object in this augmenting path by the maximum amount that will not violate $\e$-CS, and go to the next iteration.}\texshopboxnt{\pn 
\nitem{$\bullet$}  If ${\cal M}(i)$ is empty, let ${\cal O}(i)$ denote the set of objects that are assigned to some coalition partner of $i$. Raise the prices of the objects in ${\cal O}(i)$ by the maximum common amount for which the $\e$-zone of every person $i'$ in ${\cal C}(i)$ is a subset of the $\e$-zone of the same person $i'$ after the price rise.
Let $\ol {\cal O}(i)$ denote the set of objects $j\notin{\cal O}(i)$, which following the price rise, belong to the $\e$-zone of a person in ${\cal C}(i)$. Select an object $\bar j\in \ol {\cal O}(i)$, with preference given to unassigned objects. Let $(i,i_1,\ldots,i_k)$ be an alternating path such that $\bar j$ is in the $\e$-zone of person $i_k$ following the price rise of the coalition objects ${\cal O}(i)$. Let also $j_1,\ldots,j_k$ be the objects that are assigned to the persons $i_1,\ldots,i_k$ in the current assignment  ${\cal A}$. Then change ${\cal A}$ by assigning $i$ to $j_1$, $i_m$ to $j_{m+1}$ for $m=1,\ldots,k-1$, and $i_k$ to $\bar j$; any person  assigned to $\bar j$ under ${\cal A}$ becomes unassigned.   Finally, raise the price of $\bar j$ by the maximum amount that will not violate $\e$-CS, and go to the next iteration.}

The person reassignments in the preceding iteration can be viewed as a collective bid, which aims to  acquire a new object for the coalition ${\cal C}(i)$, at the expense of deassigning a person from outside the coalition. The reassignments bring the iteration closer in spirit to the aggressive auction algorithm. In particular, it can be seen that {\it if ${\cal Z}(i)$ consists of a single object (assigned or unassigned), the preceding iteration behaves identically with the aggressive auction iteration\/}. On the other hand, we should also note that person reassignments do not allow the use of coalition expansions.

\vskip-0.5pc
\subsubsection{$\e$-Scaling Variations}
\vskip-0.5pc
\pn The use of $\e$-scaling may or may not be necessary for the cooperative auction algorithm of Section 3 and its variants. After all,  with $\e=0$ the cooperative algorithm is known to be reliable and to perform well for many problems, particularly those involving a dense assignment graph (this has been established by a number of studies starting with the original paper  [Ber81]). On the other hand, $\e$-scaling may be needed to improve the robustness and the performance of both the aggressive and the cooperative  algorithms for the case of a sparse assignment graph. 

A critical step in $\e$-scaling is when a complete assignment is obtained with some value of $\e$ and then, to run the algorithm with a smaller value $\bar\e<\e$, one must discard from the assignment those pairs that do not satisfy $\bar\e$-CS. An alternative possibility is to use a variant of the auction algorithm that does not require that the initial price and assignment satisfy $\bar\e$-CS. In this variant, we try to execute the cooperative and noncooperative iterations as if $\bar\e$-CS were satisfied, and when assigned pairs $(i,j)$ not satisfying $\bar\e$-CS are encountered, to discard these pairs from the assignment as needed, while making sure that all newly assigned pairs satisfy $\bar\e$-CS. With this somewhat speculative mode of operation, progress can be made towards satisfying $\bar\e$-CS as the algorithm is running, with potentially significant computational savings. 
Such variations of $\e$-scaling may be helpful in network optimization problems where finding initial conditions that satisfy $\e$-CS requires an expensive computation.

In a related context, which is very common in practice, assignment problems are solved repeatedly with small variations in the problem's data (such as small changes in the problem's graph or values). Then there is much to be gained by reusing information in the form of prices and assignment pairs, even if they do not satisfy $\e$-CS.  
As an example, the author's paper [Ber22] has introduced auction algorithms for path construction and shortest path problems, where the initial conditions need not satisfy $\e$-CS, but are progressively rectified in the course of the algorithm. This is particularly convenient in on-line applications where the problem data changes and maintaining $\e$-CS at all times is difficult (a knowledge graph context of this type is considered by Agarwal, Bertsekas, and Liu [ABL23]). The ideas of the paper [Ber22] (and related ideas from an earlier max-flow paper by the author [Ber95a]) can be extended to the algorithms of the present paper for solving assignment problems as well as other network optimization problems. 

\vskip-0.5pc
\subsubsection{Adaptive $\e$-Scaling}
\vskip-0.5pc
\pn One possibility to improve the performance of $\e$-scaling schemes is to introduce adaptivity, whereby the value of $\e$ is modified in the course of the algorithm, depending on algorithmic progress. In particular, we may start with a small value of $\e$ and suitably increase it if some heuristic criterion suggests that a price war is underway (a simple heuristic of this type is implemented in the author's FORTRAN codes noted earlier). 

Another possibility is to {\it use a person-dependent value of $\e$\/}, so each person has his/her own value that determines the size of his/her $\e$-zone. In particular, if  the parameter value $\e_i$ is used by person $i$, we may increase $\e_i$ by some factor (up to some upper bound), each time $i$ submits an aggressive auction single-person bid, thereby expanding the $\e$-zone ${\cal Z}(i)$.  This enhances the cooperative character of iterations that involve repeat bidders, such as the ones participating in a price war.  Intuitively, in this form of adaptive $\e$-scaling, {\it a person $i$ that submits an aggressive bid repeatedly, only to be outbid later by some other person, seeks coalition partners by increasing $\e_i$ in order to get through a price war more quickly\/}.

\vskip0pc
\subsubsection{Reverse Iterations, Third Best Value, Similar Persons and Objects}
\vskip-0.5pc
\pn We note that other implementation variants of the auction algorithm have been proposed in the literature, and can be adapted to the cooperative framework of this paper, and its extensions to other network optimization problems. These include the use of reverse iterations (see Bertsekas, Casta\~non, and Tsaknakis [BCT93], the books [Ber91a], Section 4.2, and [Ber98], Section 7.2), and the ``third best" value heuristic (see Exercise 1.7, Section 4.1 of the book [Ber91], or Exercise 7.7 of the book [Ber98]). Both of these variations can be very effective, require minimal additional overhead, and have been  implemented in the author's FORTRAN codes. 

Some variations that are important from both the algorithmic and the theoretical/conceptual point of view deal with problems where there many ``similar" persons and objects [many persons $i$ with identical object sets $A(i)$ and values $a_{ij}$, $j\in A(i)$]. Problems of this type are particularly susceptible to price wars; see the books [Ber91a], Section 4.2, [Ber98], Chapter 7. The paper by Bertsekas and Casta\~non [BeC89], and the more recent papers by Walsh and Dieci [WaD17], [WaD19] propose related auction algorithms in the context of transportation problems, which can be converted into assignment problems with many similar persons and objects. Walsh has also written publicly available auction codes for transportation problems; see https://github.com/jdwalsh03/auction. Alternatively, transportation problems may  be viewed as special cases of linear single commodity network problems, and they can be addressed by corresponding natural extensions of auction algorithms. 
 
\vskip-0.5pc
\subsubsection{Special Choices of Unassigned Persons}
\vskip-0.5pc
\pn All the algorithms that we have discussed, except for the ones involving expanding coalitions, leave open the choice of the unassigned person $i$ that initiates the auction iteration. However, problems with special structure may lend themselves to special/favorable choices of $i$. For example in assignment problems that have a path construction structure, such as shortest path-type or max-flow-type  problems, it may be beneficial to choose unassigned persons in a sequence that corresponds to a candidate solution path or candidate augmenting path; see the author's paper [Ber22] for related auction algorithmic ideas.

In the context of the assignment problem, a special choice of this type corresponds to choosing the person $i$ that starts an auction iteration to be one that has just lost his/her assigned object due to an aggressive bid by another person. We will not go into further details, and instead refer to the papers [Ber91b], [Ber95a], [Ber95b], [Ber22], and the books [Ber91a], [Ber98] for discussion of such possibilities and the intuition behind them. 

\vskip0pc
\subsubsection{Heuristic Criteria for Switching to Cooperative Auction}
\vskip-0.5pc
\pn An issue that arises in combinations of conservative/aggressive and cooperative auction is to control the switch from one type of auction to another. One possibility is to forgo the aggressive iteration and do a cooperative iteration instead, if some heuristic criterion suggests that a price war is underway; for example, a relatively large number of aggressive iterations that do not produce an augmentation. This is similar to what is done in two-phase auction algorithms with $\e=0$, which start as single-person/conservative auction and switch to a cooperative auction if price wars persist, e.g., the algorithms of [Ber81] and [JoV87].

\vskip-0.5pc
\subsubsection{Dealing with Infeasibility}
\vskip-0.5pc
\pn Let us consider the case of an infeasible  problem, where there does not exist a complete assignment. In this case, the auction algorithm cannot possibly terminate. It will keep on increasing the prices of some objects by increments of at least $\e$. Furthermore, some persons will be submitting bids infinitely often, and the corresponding profits will be decreasing toward  $-\infty$. Methods  to detect infeasibility of a given problem have been developed and have been discussed in several of the author's works; see for example [Ber91a], [Ber92], [Ber98]. These methods can be easily incorporated into the algorithmic framework of this paper. 

A simple method to deal with infeasibility is to convert the problem to an equivalent feasible problem by adding a set of artificial person-object pairs to the original set of pairs. The values of these pairs should be very small, so that none of them participates in an optimal assignment unless the problem is infeasible. We refer to Section 3.3 of the tutorial paper [Ber92] for further discussion.
An alternative possibility is to first check for feasibility of the problem (before attempting to solve it) by using a low complexity bipartite matching algorithm for infeasibility detection.

Finally, let us note that if the expanding coalitions variant is used, the detection of infeasibility is simple: the problem is infeasible if and only if in the course of some cooperative iteration (with coalition expansion) we encounter an empty set of border nodes.

\vskip-1pc
\section{Concluding Remarks}
\vskip-0.5pc
\pn We have introduced a new cooperative auction iteration, and variations thereof, for symmetric linear assignment problems, which may use a positive value of $\e$, and can resolve competitive impasses and price wars without requiring the use of $\e$-scaling (although it can be used in conjunction with $\e$-scaling). The iteration is recommended when the $\e$-zone of the starting unassigned person involves multiple assigned objects, an indication of the possibility of a price war; otherwise the classical aggressive form of the auction iteration is typically preferable. The variant of the cooperative auction iteration that involves person reassignments actually coincides with the aggressive auction iteration when the $\e$-zone the starting person consists of a single object. 

The auction iterations described in this paper admit extensions to other classical network optimization problems such as asymmetric assignment, multiassignment, shortest path, $k$-shortest path, max-flow, and transportation  problems. All of these problems can in turn be viewed as special cases of the general single commodity linear network flow problem, which is commonly referred to as the {\it minimum cost flow problem} (MCNF for short) in the literature. 

We plan to discuss extensions of the cooperative auction algorithm and its variants to other network flow problems in future publications. However, it is worth mentioning here some connections between the assignment algorithms of the present paper and algorithms for the MCNF problem, which point the way to future work: 

\nitem{(a)} Conservative auction, when generalized to the MCNF problem, becomes the single node relaxation method described in Section 6.3 of the book [Ber98].
     
\nitem{(b)} Aggressive auction, when generalized to the MCNF problem, becomes the $\e$-relaxation method first proposed by the author in the paper [Ber86], and described and analyzed in detail in the books [BeT89] (Sections 5.3, 5.4), [Ber91] (Section 4.5), and [Ber98] (Section 7.4). This method is also closely related to preflow-push methods, as noted earlier.

\nitem{(c)} The variant of the combined cooperative/conservative auction algorithm ($\e=0$), which does not involve coalition expansions, when generalized to the MCNF problem, becomes the relaxation method of the paper [BeT88], and the books [Ber91] (Section  3.3) and [Ber98] (Section 6.3). 

\nitem{(d)} The purely cooperative auction algorithm with $\e=0$ and coalition expansions, when generalized to the MCNF problem, becomes the classical primal-dual (sequential shortest path) method; see [Ber98] (Section 6.2). 

\nitem{(e)} The variant of the cooperative auction algorithm that was first presented in Section 4 (person reassignments along an alternating path), when generalized to the MCNF problem with $\e=0$, becomes a variant of the relaxation method described in the paper [Ber95b] under the name ``early flow augmentations."

\nitem{(f)} An auction algorithm for the max-flow problem, given by the author in the paper [Ber95a], combines several of the variations of aggressive and cooperative auction algorithms that we have discussed. Of course, the max-flow problem has special structure (such as zero arc costs and hence no need for $\e$-scaling), which can be exploited when specializing the algorithms of the present paper to its context.

\nitem{(g)} Cooperative auction with $\e>0$, and its variants with and without coalition expansions and  person reassignments, are new algorithms, which generalize without much difficulty to the MCNF problem and its special cases noted earlier. Early ideas in this regard can be found in the paper by Bertsekas  and Casta\~ non [BeC93c], and the book [Ber98], Section 9.6.
\smskip

Another form of extension to a MCNF problem that involves a convex (rather than linear) separable cost function, is also possible. It can be based on related problem transformation ideas (see the papers by Bertsekas, Polymenakos, and Tseng [BPT97], [BPT98], and the textbook [Ber98], Chapter 9).

A basic mechanism for extension of auction algorithms to  MCNF problems and  special cases thereof is to first convert such problems to assignment problems, by using well known transformations, then apply one of the algorithms of the present paper, and then streamline the computations for efficiency. 
However, as a practical matter one should not try to literally convert one of the assignment algorithms of the present paper to a new problem structure. Instead one should aim to combine and adapt the principal algorithmic ideas presented in this paper, in sensible ways that experimentally can be shown to work well for the given type of problem. These ideas are conservative, aggressive, and cooperative price rises and augmentations, under the umbrella of the mathematically fundamental approximation framework of  $\e$-CS, and the intuitive framework of auction-based economic competition.

Let us also mention extensions of the purely cooperative auction iteration (possibly in combination with aggressive auction iterations) that allow multiple unassigned persons to jointly initiate an iteration. This type of extension is not discussed in the present paper, but may be relevant, among others, to distributed auction algorithms involving multiple bidders submitting bids in parallel. See the book [BeT89], Sections 5.3 and 6.5, for related discussions of distributed asynchronous aggressive auction algorithms, and also the papers by Bertsekas and Casta\~ non [BeC93a], [BeC93b] for distributed asynchronous implementations of the Hungarian method and primal-dual methods.

We note that beyond their use in addressing the MCNF problem, our algorithmic ideas lend themselves well for incorporation in heuristics for assignment-like problems, which are more difficult than the linear assignment problem that we have considered in this paper. Such problems include multi-dimensional assignment, combinatorial auctions, dynamic task allocation, and multiagent/multi-robot  problems. A noteworthy context is to use an auction algorithm as a base heuristic for a rollout algorithm; see the books [Ber98] (Section  10.5), [Ber20a] (Section 3.4.2), and [Ber23b] (Chapter 2).

Finally, let us mention an interesting connection with reinforcement learning. One of the important favorable characteristics of auction algorithms is that the final prices obtained from solution of a given assignment problem can be used as initial prices for applying the algorithms to other problems, which are structurally similar. This suggests that one may try to ``learn" favorable initial prices from data and encode this knowledge into a neural network that can supply on demand good initial prices for a given problem. Work on machine learning and neural network approaches towards assignment problems is at a very early stage at present; see e.g., Lee et al.\ [LXY18], Emami et al.\ [EPE20], and Aironi, Cornell, and Squartini [ACS22]. It is reasonable to expect that auction algorithms and their intuitive economic competition-like mechanism lend themselves well to this line of research.

\vskip-1pc
\vfill\eject
\section{References}
\vskip-0.5pc

\ref[ABL23] Agrawal, G., Bertsekas, D., Liu, H., 2023.\ ``Auction-Based Learning for Question Answering over Knowledge Graphs," Information, Vol.\ 14, 336, https://doi.org/10.3390/ info14060336.

\ref[ACS22]  Aironi, C., Cornell, S., and Squartini, S., 2022.\ ``Tackling the Linear Sum Assignment Problem with Graph Neural Networks," in International Conference on Applied Intelligence and Informatics, Springer, pp.\ 90-101.

\ref[AMO88] Ahuja, R.\ K., Magnanti, T.\ L., and Orlin, J.\ B., 1988.\ Network Flows, dspace.mit.edu.

\ref[AMO89] Ahuja, R.\ K., Magnanti, T.\ L., and Orlin, J.\ B., 1989.\ ``Network Flows," in Handbooks in Operations Research and Management Science, Vol.\ 1, Optimization, Nemhauser, G.\ L., et al.\ (eds.), North-Holland, Amsterdam, pp. 211-369.

\ref[APP22] Aziz, H., Pal, A., Pourmiri, A., Ramezani, F., and Sims, B., 2022.\ ``Task Allocation Using a Team of Robots," Current Robotics Reports, Vol.\ 3, pp.\ 227-238.

\ref[APV22] Alfaro, C.\ A., Perez, S.\ L., Valencia, C.\ E., and Vargas, M.\ C., 2022.\ ``The Assignment Problem Revisited," Optimization Letters, Vol.\ 16, pp.\ 1531-1548.

\ref[AhO89] Ahuja, R.\ K., and Orlin, J.\ B., 1989.\ ``A Fast and Simple Algorithm for the Maximum Flow Problem," Operations Research, Vol.\ 37, pp. 748-759.

\ref [Ami94] Amini, M.\ M., 1994.\ ``Vectorization of an Auction Algorithm for Linear Cost Assignment
Problem," Comput.\ Ind.\ Eng., Vol.\ 26, pp.\ 141-149.

\old{
\ref[BBS95] Barto, A.\ G., Bradtke, S.\ J., and Singh, S.\ P., 1995.
``Learning to Act Using Real-Time Dynamic Programming," Artificial Intelligence, 
Vol.\ 72, pp.\ 81-138.
}

\ref[BCE95] Bertsekas, D.\ P., Casta\~ non, D.\ A., Eckstein, J., and Zenios, S.,
1995.\ ``Parallel Computing in Network Optimization," Handbooks in OR and
MS,  Ball, M.\ O., Magnanti, T.\ L., Monma, C.\ L., and Nemhauser, G.\ L.\
(eds.),  Vol.\ 7, North-Holland, Amsterdam, pp.\ 331-399.

\ref[BCT93] 
Bertsekas, D.\ P., Casta\~ non, D.\ A., and Tsaknakis, H.,
1993.\ ``Reverse Auction and the Solution of Inequality Constrained Assignment
Problems," SIAM J.\ on Optimization, Vol.\ 3, pp.\ 268-299.

\ref[BDM12] Burkard, R., Dell' Amico, M., and Martello, S., 2012.\ Assignment Problems, Society for Industrial and Applied Mathematics.

\ref [BFH03] Brenier, Y., Frisch, U., Henon, M., Loeper, G., Matarrese, S., Mohayaee, R., and Sobolevskii, A., 2003.\ ``Reconstruction of the Early Universe as a Convex Optimization Problem," Monthly Notices of the Royal Astronomical Society, Vol.\ 346, pp.\ 501-524.

\ref[BGM97] Beraldi, P., Guerriero, F., and Musmanno, R., 1997.\ ``Efficient Parallel Algorithms for the Minimum Cost Flow Problem," Journal of Optimization Theory and Applications, Vol.\ 95, pp.\ 501-530.

\ref[BGM01] Beraldi, P., Guerriero, F., and Musmanno, R., 2001.\ ``Parallel Algorithms for Solving the Convex Minimum Cost Flow Problem," Computational Optimization and Applications, Vol.\ 18, pp.\ 175-190.

\old{
\ref [BPS95] Bertsekas, D.\ P., Pallottino, S., and Scutell\`a, M.\ G., 1995.\ ``Polynomial
Auction Algorithms for Shortest Paths,'' Computational Optimization and Applications,
Vol.\ 4, pp.\ 99-125.
}

\ref[BPS07] Bayati, M., Prabhakar, B., Shah, D., and Sharma, M., 2007.\ ``Iterative Scheduling Algorithms," in IEEE INFOCOM 2007-26th IEEE International Conf.\ on Computer Communications, pp.\ 445-453.

\ref [BPT97] Bertsekas, D.\ P., Polymenakos, L.\ C., and Tseng, P., 1997.\ ``An Epsilon-Relaxation Method for Convex Network Optimization Problems," SIAM J.\ on Optimization, Vol.\ 7, pp.\ 853-870.

\ref [BPT98] Bertsekas, D.\ P., Polymenakos, L.\ C., and Tseng, P., 1998.\ 
``Epsilon-Relaxation and Auction Methods for Separable Convex Cost Network Flow Problems," in Network Optimization, by P.\ M.\ Pardalos, D.\ W.\ Hearn, and W.\ W.\ Hager (eds.), Lecture Notes in Economics and Mathematical Systems, Springer-Verlag, N.Y., pp.\ 103-126.

\ref[BSS08] Bayati, M., Shah, D., and Sharma, M., 2008.\ ``Max-Product for Maximum Weight Matching: Convergence, Correctness, and LP Duality," IEEE Trans.\ on Information Theory, Vol.\ 54, pp.\ 1241-1251.

\ref [BaF88] Bar-Shalom, Y., and Fortman, T.\ E., 1988.\ Tracking and Data
Association, Academic Press, N.\ Y.

\ref [Bar90] Bar-Shalom, Y., 1990.\ Multitarget-Multisensor Tracking: Advanced Applications, Artech House, Norwood, MA.

\ref[BeC89] Bertsekas, D.\ P., and Casta\~ non, D.\ A., 1989.\ ``The Auction Algorithm for the Transportation Problem," Annals of Operations Research, Vol.\ 20, pp.\ 67-96.

\ref[BeC91] Bertsekas, D.\ P., and Casta\~ non, D.\ A., 1991.\ ``Parallel Synchronous and Asynchronous Implementations of the Auction Algorithm," Parallel Computing, Vol.\ 17, pp.\ 707-732.

\ref[BeC93a] Bertsekas, D.\ P., and Casta\~ non, D.\ A., 1993.\ ``Parallel Asynchronous Hungarian Methods for the Assignment Problem," ORSA J.\ on Computing, Vol.\ 5, pp. 261-274.

\ref[BeC93b] Bertsekas, D.\ P., and Casta\~ non, D.\ A., 1993.\ ``Parallel Primal-Dual Methods for the Minimum Cost Flow Problem," Computational Optimization and Applications, Vol.\ 2, pp. 317-336.

\ref[BeC93c] Bertsekas, D.\ P., and Casta\~ non, D.\ A., 1993.\ ``A Generic Auction Algorithm
for the Minimum Cost Network Flow Problem," Computational Optimization and Applications,
Vol.\ 2, pp.\ 229-260.

\ref[BeE87] Bertsekas, D.\ P., and Eckstein, J., 1987.\ ``Distributed Asynchronous Relaxation Methods for Linear Network Flow Problems," IFAC Proceedings, Vol.\ 20, pp.\ 103-114.

\ref[BeE88] Bertsekas, D.\ P., and Eckstein, J., 1988.\ ``Dual Coordinate Step
Methods for Linear Network Flow Problems,'' Math.\ Programming,
Series B, Vol.\ 42, pp.\ 203-243.

\ref[BeG97] Beraldi, P., and Guerriero, F., 1997.\ ``A Parallel Asynchronous Implementation of the $\e$-Relaxation Method for the Linear Minimum Cost Flow Problem," Parallel Computing, Vol.\ 23, pp.\ 1021-1044.

\ref[BeM19] Bertozzi, A.\ L., and Merkurjev, E., 2019.\ ``Graph-Based Optimization Approaches for Machine Learning, Uncertainty Quantification and Networks," in Handbook of Numerical Analysis, Vol.\ 20, pp 503-531.

\ref[BeT88] Bertsekas, D.\ P., and Tseng, P., 1988.\ ``Relaxation Methods for
Minimum Cost Ordinary and Generalized Network Flow Problems,'' 
Operations Research, Vol.\ 36, pp.\ 93-114.

\ref [BeT89]  Bertsekas, D.\ P., and Tsitsiklis, J.\ N., 1989.\ Parallel and
Distributed Computation: Numerical Methods, Prentice-Hall, Engl.\ 
Cliffs, N.\ J.\  (can be  downloaded from the author's website).

\ref[BeT94] Bertsekas, D.\ P., and Tseng, P., 1994.\ ``RELAX-IV: A Faster Version of the RELAX Code for Solving Minimum Cost Flow Problems," Report LIDS-P-2276, MIT.

\ref [BeT97]  
 Bertsimas, D., and Tsitsiklis, J.\ N., 1997.\ Introduction to Linear Optimization,
Athena Scientific, Belmont, MA.

\ref [Ber79] Bertsekas, D.\ P., 1979.\ ``A Distributed Algorithm for the Assignment Problem," Lab.\ for Information and Decision Systems Report, MIT, May 1979.

\ref [Ber81] Bertsekas, D.\ P., 1981.\ ``A New Algorithm for the Assignment
Problem,'' Math.\ Programming, Vol.\ 21, pp.\ 152-171.

\ref [Ber85] Bertsekas, D.\ P., 1985.\ ``A Unified Framework for Minimum Cost
Network Flow Problems,'' Math.\ Programming,
Vol.\ 32, pp.\ 125-145.

\ref [Ber86] Bertsekas, D.\ P., 1986.\ ``Distributed Asynchronous
Relaxation Methods for Linear Network Flow Problems,'' 
1986 25th IEEE Conference on Decision and Control, pp.\ 2101-2106.

\ref [Ber88] Bertsekas, D.\ P., 1988.\ ``The Auction Algorithm: A Distributed Relaxation Method for the Assignment Problem," Annals of Operations Research, Vol.\ 14, pp.\ 105-123.

\ref[Ber90] Bertsekas, D.\ P., 1990.\ ``The Auction Algorithm for Assignment and
Other Network Flow Problems: A Tutorial," Interfaces, Vol.\ 20, pp.\
133-149.

\ref [Ber91a] Bertsekas, D.\ P., 1991.\ Linear Network Optimization, MIT Press, Cambridge, MA.

\ref [Ber91b] Bertsekas, D.\ P., 1991.\ ``An Auction Algorithm for Shortest Paths," 
SIAM J.\ on Optimization, Vol.\ 1, pp.\ 425-447.

\ref [Ber92] Bertsekas, D.\ P., 1992.\ ``Auction Algorithms for Network Flow Problems: A Tutorial Introduction," Computational Optimization and Applications, Vol.\ 1, pp.\ 7-66. 

\ref [Ber93] Bertsekas, D.\ P., 1993.\ ``Mathematical Equivalence of the Auction Algorithm for Assignment and the Epsilon-Relaxation (Preflow-Push) Method for Min Cost Flow," in Large Scale Optimization, Hager W.\ W., Hearn D.\ W., Pardalos P.\ M. (eds), Springer, Boston, MA. 

\ref [Ber95a] Bertsekas, D.\ P., 1995.\ ``An Auction Algorithm for the Max-Flow Problem,"
J.\ of Optimization Theory and Applications, Vol.\ 87, pp.\ 69-101.

\ref [Ber95b] Bertsekas, D.\ P., 1995.\ ``An Auction/Sequential Shortest Path Algorithm for the Minimum Cost Network Flow Problem," Report LIDS-P-2146, MIT.

\ref [Ber98] Bertsekas, D.\ P., 1998.\
Network Optimization: Continuous and Discrete Models, Athena Scientific,
Belmont, MA (can be downloaded from the author's website).

\old{
\ref[Ber17] Bertsekas, D.\ P., 2017.\ Dynamic Programming and Optimal Control, Vol.\ I, Athena Scientific, Belmont, MA.
}

\old{
\ref[Ber19] Bertsekas, D.\ P., 2019.\ Reinforcement Learning and Optimal Control, Athena Scientific, Belmont, MA.
}

\ref[Ber20a] Bertsekas, D.\ P., 2020.\
Rollout, Policy Iteration, and Distributed Reinforcement Learning, Athena Scientific, Belmont, MA.

\ref[Ber20b] Bertsekas, D.\ P., 2020.\ ``Constrained Multiagent Rollout and Multidimensional Assignment with the Auction Algorithm," arXiv preprint, arXiv:2002.07407.

\ref [Ber22] Bertsekas, D.\ P., 2022.\
``Auction Algorithms for Path Planning, Network Transport, and Reinforcement Learning," Arizona State University/SCAI Report; arXiv:22207.09588.

\ref[Ber23a] Bernard, F., 2023.\ ``Fast Linear Assignment Problem using Auction Algorithm," MATLAB Central File Exchange," https://www.mathworks.com/matlabcentral/fileexchange/48448-fast-linear-assignment-problem-using-auction-algorithm-mex.

\ref [Ber23b] Bertsekas, D.\ P., 2023.\ A Course in Reinforcement Learning, Athena Scientific, Belmont, MA. 

\old{
\ref [Ber22] Bertsekas, D.\ P., 2022.\
Auction Algorithms for Assignment, Path Planning, and Network Transport, Athena Scientific,
Belmont, MA (in preparation).
}

\old{
\ref[Ber22b] Bertsekas, D.\ P., 2022.\
Lessons from AlphaZero for Optimal, Model Predictive, and Adaptive Control, Athena Scientific,
Belmont, MA  \{also available as an ebook from Google Books, and on-line from the author's website).
}

\old{
\ref[BiT22] Bicciato, A., and Torsello, A., 2022.\ ``GAMS: Graph Augmentation with Module Swapping," Proc.\ of ICPRAM, pp.\ 249-255.
}

\ref[Bla86] Blackman, S.\ S., 1986.\ Multi-Target Tracking with Radar Applications, Artech House, Dehdam, MA.

\ref[BuC99] Burkard, R.\ E., and Cela, E., 1999.\ ``Linear Assignment Problems and Extensions," in Handbook of Combinatorial Optimization: Springer, Boston, Supplement Vol.\ A, pp.\ 75-149.

\ref[BuT09] Bus, L., and Tvrdík, P., 2009. ``Towards Auction Algorithms for Large Dense Assignment Problems," Computational Optimization and Applications, Vol.\ 43, pp.\ 411-436.

\ref[CBH09] Choi, H.\ L., Brunet, L., and How, J.\ P., 2009.\ ``Consensus-Based Decentralized Auctions for Robust Task Allocation," IEEE Transactions on Robotics, Vol.\ 25, pp.\ 912-926.

\old{
\ref[CLG22] Clark, A., de Las Casas, D., Guy, A., Mensch, A., Paganini, M., Hoffmann, J., Damoc, B., Hechtman, B., Cai, T., Borgeaud, S., and Van Den Driessche, G.\ B., 2022. ``Unified Scaling Laws for Routed Language Models," Proc.\ International Conference on Machine Learning, pp.\ 4057-4086.
}

\ref[Cas92] Casta\~ non, D.\ A., 1992.\ ``New Assignment Algorithms for Data Association," in Signal and Data Processing of Small Targets, Vol.\ 1698, pp.\ 313-323.

\ref[Cas93] Casta\~ non, D.\ A., 1993. ``Reverse Auction Algorithms for Assignment
Problems," in Algorithms for Network Flows and Matching," Johnson, D.\ S., and
McGeoch, C.\ C., (eds.), American Math.\ Soc., Providence, RI, pp. 407-429.

\ref[CeZ97] Censor, Y., and Zenios, S.\ A., 1997.\ Parallel Optimization: Theory, Algorithms, and Applications, Oxford University Press.

\ref[ChM89] Cheriyan, J., and Maheshwari, S.\ N., 1989.\ ``Analysis of Preflow Push
Algorithms for Maximum Network Flow,'' SIAM J.\ Computing, Vol.\ 18, pp.\
1057-1086.

\ref[DLT19] Duan, X., Liu, H., Tang, H., Cai, Q., Zhang, F., and Han, X., 2019.\ ``A Novel Hybrid Auction Algorithm for Multi-UAVs Dynamic Task Assignment," IEEE Access, 8, pp.\ 86207-86222.

\old{
\ref[DeM89] Derigs, U., and Meier, W., 1989.\ ``Implementing Goldberg's Max-Flow Algorithm - A
Computational Investigation," Zeitschrif fur Operations Research, Vol.\ 33, pp.\ 383-403.
}

\ref[DeV03] De Vries, S., and Vohra, R.\ V., 2003.\ ``Combinatorial Auctions: A Survey," INFORMS Journal on Computing, Vol.\ 15, pp.\ 284-309.

\ref[EPE20] Emami, P., Pardalos, P.\ M., Elefteriadou, L., and Ranka, S., 2020.\ ``Machine Learning Methods for Data Association in Multi-Object Tracking," ACM Computing Surveys (CSUR), Vol.\ 53, pp.\ 1-34.

\ref[FrS06] Frisch, U., and Sobolevskii, A., 2006.\ ``Application of Optimal Transport Theory to Reconstruction of the Early Universe," Journal of Mathematical Sciences, Vol.\ 133, pp.\ 1539-1542.

\ref[GBG23] Garces, D., Bhattacharya, S., Gil, S., and Bertsekas, D.\ P., 2023.\ ``Multiagent Reinforcement Learning for Autonomous Routing and Pickup Problem with Adaptation to Variable Demand," 2023 IEEE International Conference on Robotics and Automation (ICRA), pp.\ 3524-3531.

\ref [Gal16] Galichon, A., 2016.\ Optimal Transport Methods in Economics, Princeton University Press.

\ref[GoT86] Goldberg, A.\ V., and Tarjan, R.\ E., 1986.\ ``A New Approach to
the Maximum Flow Problem,'' Proc.\ 18th ACM STOC, pp.\ 136-146.

\ref[GoT90] Goldberg, A.\ V., and Tarjan, R.\ E., 1990.\ ``Solving Minimum
Cost Flow Problems by Successive Approximation,'' Math.\ of Operations Research,
Vol.\ 15, pp.\ 430-466.

\ref[HZX19] Huang, Y., Zhang, Y., and Xiao, H., 2019.\ ``Multi-Robot System Task Allocation Mechanism for Smart Factory," in 2019 IEEE 8th Joint International Information Technology and Artificial Intelligence Conference (ITAIC), pp.\ 587-591.

\ref[JME18] Jacobs, M., Merkurjev, E., and Esedoglu, S., 2018.\ ``Auction Dynamics: A Volume Constrained MBO Scheme," Journal of Computational Physics, Vol.\ 354, pp.\ 288-310.

\ref[JoV87] Jonker, R., and Volgenant, T., 1987.\ ``A Shortest Augmenting Path Algorithm for Dense and Sparse Linear Assignment Problems," Computing, Vol.\ 38.

\ref[KSS14] Kollias, G., Sathe, M., Schenk, O., and Grama, A., 2014.\ ``Fast Parallel Algorithms for Graph Similarity and Matching," Journal of Parallel and Distributed Computing, Vol.\ 74, pp.\ 2400-2410.

\ref[Kar74] Karzanov, A.\ V., 1974.\ ``Determining the Maximal Flow in a
Network with the Method of Preflows,'' Soviet Math Dokl., Vol.\ 15, pp.\
1277-1280.

\ref[KhA21] Khosla, M., and Anand, A., 2021.\ ``Revisiting the Auction Algorithm for Weighted Bipartite Perfect Matchings," arXiv preprint arXiv:2101.07155.

\ref[KoY94] Kosowsky, J.\ J., and Yuille, A.\ L., 1994.\ ``The Invisible Hand Algorithm: Solving the Assignment Problem with Statistical Physics," Neural Networks, Vol.\ 7, pp.\ 477-490.

\ref[Kuh55] Kuhn, H.\ W., 1955.\ ``The Hungarian Method for the Assignment
Problem,'' Naval Research Logistics Quarterly, Vol.\ 2, pp.\ 83-97. 

\ref[LCS14] Luo, L., Chakraborty, N., and Sycara, K., 2014.\ ``Provably-Good Distributed Algorithm for Constrained Multi-Robot Task Assignment for Grouped Tasks," IEEE Transactions on Robotics, Vol.\ 31, pp.\ 19-30.

\ref[LGO20] Lujak, M., Giordani, S., Omicini, A., and Ossowski, S., 2020.\ ``Decentralizing Coordination in Open Vehicle Fleets for Scalable and Dynamic Task Allocation," Complexity, pp.\ 1-21.

\ref[LMH21] Levy, B., Mohayaee, R., and von Hausegger, S., 2021.\ ``A Fast Semidiscrete Optimal Transport Algorithm for a Unique Reconstruction of the Early Universe," Monthly Notices of the Royal Astronomical Society, Vol.\ 506, pp.\ 1165-1185.

\ref[LXY18] Lee, M., Xiong, Y., Yu, G., Li, G.\ Y., 2018.\ ``Deep Neural Networks for Linear Sum Assignment Problems," IEEE Wirel.\ Commun.\ Lett., Vol.\ 7, pp.\ 962-965.

\ref[LZX23] Li, H., Zhu, H., Xu, D., Lin, X., Jiao, G., Song, Y., and Huang, M., 2023.\ ``Dynamic Task Allocation Based on Auction in Robotic Mobile Fulfilment System," Journal of Industrial and Management Optimization, Vol.\ 19.

\ref[Lav08] Lavaux, G., 2008.\ ``Lagrangian Reconstruction of Cosmic Velocity Fields," Physica D: Nonlinear Phenomena, Vol.\ 237, pp.\ 2139-2144.

\ref[LiS13] Liu, L., and Shell, D.\ A., 2013.\ ``Optimal Market-Based Multi-Robot Task Allocation via Strategic Pricing," Robotics: Science and Systems, Vol.\ 9, pp.\ 33-40.

\ref[LuM20] Lujak, M., and Matezovic, M., 2020.\ ``On Efficiency in Dynamic Multi-Robot Task Allocation," in AIRO Proceedings, pp.\ 49-53.

\ref[Mal97] Malkoff, D.\ B., 1997.\ ``Evaluation of the Jonker-Volgenant-Casta\~ non (JVC) Assignment Algorithm for Track Association", Proc.\ SPIE 3068, Signal Processing, Sensor Fusion, and Target Recognition; https://doi.org/10.1117/12.280801.

\ref[MBM19] Metivier, L., Brossier, R., Merigot, Q., and Oudet, E., 2019.\ ``Graph Space Optimal Transport for FWI: Auction Algorithm, Application to the 2d Valhall Case Study," in 81st EAGE Conference and Exhibition, European Association of Geoscientists and  Engineers.

\ref[MSC16] Morgan, D., Subramanian, G.\ P., Chung, S.\ J., and Hadaegh, F.\ Y., 2016.\ ``Swarm Assignment and Trajectory Optimization Using Variable-Swarm, Distributed Auction Assignment and Sequential Convex Programming," The International Journal of Robotics Research, Vo.\ 35, pp.\ 1261-1285.

\ref[MeT21] Merigot, Q., and Thibert, B., 2021.\ ``Optimal Transport: Discretization and Algorithms," in Handbook of Numerical Analysis,  Elsevier, Vol.\ 22, pp.\ 133-212.

\ref[Mer20] Merkurjev, E., 2020.\ ``A Fast Graph-Based Data Classification Method with Applications to 3D Sensory Data in the Form of Point Clouds," Pattern Recognition Letters, Vol.\ 136, pp.154-160.

\ref[NSB23] Nurlanov, Z., Schmidt, F.\ R., and Bernard, F., 2023.\ ``Universe Points Representation Learning for Partial Multi-Graph Matching," in Proc.\ of the AAAI Conference on Artificial Intelligence, Vol.\ 37, pp.\ 1984-1992.

\ref[NVJ16] Nascimento, A.\ D.\ P., Vasconcelos, C.\ N., Jamel, F.\ S., and Sena, A.\ C., 2016.\ ``A Hybrid Parallel Algorithm for the Auction Algorithm in Multicore Systems," in 2016 International Symposium on Computer Architecture and High Performance Computing Workshops (SBAC-PADW), pp.\ 73-78.

\ref[NaL16] Naparstek, O., and Leshem, A., 2016.\ ``Expected Time Complexity of the Auction Algorithm and the Push Relabel Algorithm for Maximum Bipartite Matching on Random Graphs," Random Structures and Algorithms, Vol.\ 48, pp.\ 384-395.

\ref[OKS20] Otte, M., Kuhlman, M.\ J., and Sofge, D., 2020.\ ``Auctions for Multi-Robot Task Allocation in Communication Limited Environments," Autonomous Robots, Vol.\ 44, pp.\ 547-584.

\ref[OrA92] Orlin, J.\ B., and Ahuja, R.\ K., 1992.\ ``New Scaling Algorithms for the Assignment and Minimum Mean Cycle Problems," Math.\ Programming, Vol.\ 54, pp.\ 41-56.

\ref [PBW92] Pattipati, K.\ R., Deb, S.,  Bar-Shalom, Y., and Washburn, R.\ B., 1992.\ ``A New
Relaxation Algorithm and Passive Sensor Data Association,"  IEEE Trans.\ Automatic
Control, Vol.\ 37, pp.\ 198-213.

\ref[PPB01] Popp, R.\ L., Pattipati, K.\ R., and Bar-Shalom, Y., 2001.\ ``$m$-Best SD Assignment Algorithm with Application to Multitarget Tracking," IEEE Transactions on Aerospace and Electronic Systems, Vol.\ 37, pp.\ 22-39.

\ref[PaU99] Parkes, D.\ C., and Ungar, L.\ H., 2000.\ ``Iterative Combinatorial Auctions: Theory and Practice." Aaai/iaai, 7481, p.\ 53.

\ref[PeC19] Peyre, G., and Cuturi, M., 2019.\ ``Computational Optimal Transport: With Applications to Data Science," Foundations and Trends in Machine Learning," Vol.\ 11, pp.\ 355-607.

\old{
\ref[PoB94] Polymenakos, L.\ C., and Bertsekas, D.\ P., 1994.\ ``Parallel Shortest Path Auction Algorithms," Parallel Computing, Vol.\ 20, pp.\ 1221-1247.
}

\ref [PoR97] Poore, A.\ B., and Robertson, A.\ J.\ A., 1997.\ ``New Lagrangian Relaxation Based Algorithm for a Class of Multidimensional Assignment Problems," Computational Optimization and Applications, 
Vol.\ 8, pp.\ 129-150.

\ref [Poo94] Poore, A.\ B., 1994.\ ``Multidimensional Assignment Formulation of Data Association Problems Arising from Multitarget Tracking and Multisensor Data Fusion," Computational Optimization and Applications, Vol.\ 3, pp.\ 27-57.

\ref[SSB12] Sathe, M., Schenk, O., and Burkhart, H., 2012.\ ``An Auction-Based Weighted Matching Implementation on Massively Parallel Architectures," Parallel Computing, Vol.\ 38, pp.\ 595-614.

\ref[SSN21] Sena, A.\ C., Silva, M.\ N., and Nascimento, A.\ P., 2021.\ ``An Efficient Vectorized Auction Algorithm for Many-Core and Multicore Architectures," in Latin American High Performance Computing Conference, Springer, pp.\ 76-90.

\ref[San15] Santambrogio, F., 2015.\ Optimal Transport for Applied Mathematicians, Springer Intern.\ Publ.

\ref [Sch16] Schmitzer, B., 2016.\ ``A Sparse Multiscale Algorithm for Dense Optimal Transport," J.\ of Mathematical Imaging and Vision, Vol.\  56, pp.\ 238-259.

\ref [Sch19] Schmitzer, B., 2019.\ ``Stabilized Sparse Scaling Algorithms for Entropy Regularized Transport Problems," SIAM Journal on Scientific Computing, Vol.\ 41, pp.\ A1443-A1481.

\ref[ShV82] Shiloach, Y., and Vishkin, U., 1982.\ ``An $O(n^2\log n)$ Parallel Max-Flow
Algorithm," J.\ Algorithms, Vol.\ 3, pp.\ 128-146.

\ref[TZG18] Tang, J., Zhu, K., Guo, H., Gong, C., Liao, C., and Zhang, S., 2018.\ ``Using Auction-Based Task Allocation Scheme for Simulation Optimization of Search and Rescue in Disaster Relief," Simulation Modelling Practice and Theory, Vol.\ 82, pp.\ 132-146.

\ref[TsB00] Tseng, P., and Bertsekas, D.\ P., 2000.\ ``An $\e$-Relaxation Method for Separable Convex Cost Generalized Network Flow Problems," Mathematical Programming, Vol.\ 88, pp.\ 85-104.

\ref [Vil09] Villani, C., 2009.\ Optimal Transport: Old and New, Springer, Berlin.

\ref [Vil21] Villani, C., 2021.\ Topics in Optimal Transportation, American Mathematical Society.

\ref[WLY23] Wang, Y., Li, H. and Yao, Y., 2023.\ ``An Adaptive Distributed Auction Algorithm and its Application to Multi-AUV Task Assignment," Science China Technological Sciences, pp.\ 1-10.

\ref[WMW22] Wang, C., Mei, D., Wang, Y., Yu, X., Sun, W., Wang, D., and Chen, J., 2022.\ ``Task Allocation for Multi-AUV System: A Review," Ocean Engineering, Vol.\ 266, p.\ 112911.

\ref[WaD17] Walsh, J.\ D., and Dieci, L., 2017.\ ``General Auction Method for Real-Valued Optimal Transport," arXiv preprint arXiv:1705.06379.

\ref[WaD19] Walsh, J.\ D., and Dieci, L., 2019.\ ``A Real-Valued Auction Algorithm for Optimal Transport," Statistical Analysis and Data Mining: The ASA Data Science Journal, 12(6), pp.\ 514-533.

\old{
\ref[WaX12] Wang, J., and Xia, Y., 2012.\ ``Fast Graph Construction Using Auction Algorithm," arXiv preprint arXiv:1210.4917.
}

\ref [WeZ91] Wein, J., and Zenios, S.\ A., 1991.\ ``On the Massively Parallel Solution of the
Assignment Problem,'' J.\ of Parallel and Distributed Computing, Vol.\ 13, pp.\ 228-236.

\ref[ZCH20] Zhou, J., Cui, G., Hu, S., Zhang, Z., Yang, C., Liu, Z., Wang, L., Li, C., and Sun, M., 2020.\ ``Graph Neural Networks: A Review of Methods and Applications," AI Open, Vol.\ 1, pp.\ 57-81.

\ref[ZSP08] Zavlanos, M.\ M., Spesivtsev, L., and Pappas, G.\ J., 2008.\ ``A Distributed Auction Algorithm for the Assignment Problem," Proc.\ 47th IEEE Conference on Decision and Control, pp.\ 1212-1217.

\ref[Zak95] Zaki, H.\ A., 1995.\ ``A Comparison of Two Algorithms for the Assignment Problem," Computational Optimization and Applications, Vol.\ 4, pp.\ 23-45.

\end

%% file: TEXSHOP_macros_new.tex

\def\ignore#1{}
 

\newcount\sectnum
\newcount\subsectnum
\newcount\eqnumber

\global\eqnumber=1\sectnum=0


\def\lab{(\the\sectnum.\the\eqnumber)}



\def\show#1{}



\def\smskip{\vskip 5 pt}
\def\medskip{\vskip 10 pt}
\def\bigskip{\vskip 15 pt}
\def\pn{\par\noindent}

\def\frac#1#2{{#1\over #2}}

\def\ol#1{\overline{#1}}

\def\p{\pi}

\def\e{\epsilon}

\def\tl{\tilde}

\def\old#1{}
\def\leaderfill{\leaders\hbox to 1em{\hss.\hss}\hfill}


\parindent=2pc
\baselineskip=15pt
\vsize=8.7 true in
\voffset=0.125 true in
\parskip=3pt


\def\minprob#1#2#3{$$\eqalign{&\hbox{minimize\ \ }#1\cr &\hbox{subject to\ \
}#2\cr}\ifnum 0=#3{}\else\eqno(#3)\fi$$}        
     
\def\maxprob#1#2#3{$$\eqalign{&\hbox{maximize\ \ }#1\cr &\hbox{subject to\ \
}#2\cr}\ifnum 0=#3{}\else\eqno(#3)\fi$$}        
     
\def\aligntwo#1#2#3#4#5{$$\eqalign{#1&#2\cr #3&#4\cr}
\ifnum 0=#5{}\else\eqno(#5)\fi$$}
\def\alignthree#1#2#3#4#5#6#7{$$\eqalign{#1&#2\cr #3&#4\cr #5&#6\cr}
\ifnum 0=#7{}\else\eqno(#7)\fi$$}


\def\eqnum{\eqno{\hbox{(\the\sectnum.\the\eqnumber)}\global\advance\eqnumber
by1}}

\def\eqnu{\eqno{\hbox{(\the\sectnum.\the\eqnumber)}\global\advance\eqnumber
by1}}

\newcount\examplnumber
\def\examplnum{\global\advance\examplnumber by1}

\newcount\figrnumber
\def\figrnum{\global\advance\figrnumber by1}

\newcount\propnumber
\def\propnum{\global\advance\propnumber by1}

\newcount\defnumber
\def\defnum{\global\advance\defnumber by1}

\newcount\lemmanumber
\def\lemmanum{\global\advance\lemmanumber by1}

\newcount\assumptionnumber
\def\assumptionnum{\global\advance\assumptionnumber by1}

\newcount\conditionnumber
\def\conditionnum{\global\advance\conditionnumber by1}

\def\exampl{\the\sectnum.\the\examplnumber}
\def\figr{\the\sectnum.\the\figrnumber}
\def\propn{\the\sectnum.\the\propnumber}
\def\defn{\the\sectnum.\the\defnumber}
\def\lemman{\the\sectnum.\the\lemmanumber}
\def\assumptionn{\the\sectnum.\the\assumptionnumber}
\def\condn{\the\sectnum.\the\conditionnumber}

\def\section#1{\goodbreak\vskip 3pc plus 6pt minus 3pt\leftskip=-2pc
   \global\advance\sectnum by 1\eqnumber=1
\global\examplnumber=1\figrnumber=1\propnumber=1\defnumber=1\lemmanumber=1\assumptionnumber=1 \conditionnumber =1 \subsectnum=0%
   \line{\hfuzz=1pc{\hbox to 3pc{\bf 
   \vtop{\hfuzz=1pc\hsize=38pc\hyphenpenalty=10000\noindent\uppercase{\the\sectnum.\quad #1}}\hss}}
			\hfill}
			\leftskip=0pc\nobreak\tenf
			\vskip 1pc plus 4pt minus 2pt\noindent\ignorespaces}



\def\sect#1{\noindent\leftskip=-2pc\tenf
   \goodbreak\vskip 1pc plus 4pt minus 2pt
                \global\advance\subsectnum by 1\eqnumber=1
   \line{\hfuzz=1pc{\hbox to 3pc{\bf 
   \vtop{\hfuzz=1pc\hsize=38pc\hyphenpenalty=10000\noindent\uppercase{{\bf #1}}}\hss}}
                        \hfill}
   \leftskip=0pc\nobreak\tenf
                        \vskip 1pc plus 4pt minus 2pt\nobreak\noindent\ignorespaces}

\def\subsection#1{\noindent%
   \goodbreak\vskip 1pc plus 4pt minus 2pt%
 		\global\advance\subsectnum by 1%
   \line{\hfuzz=1pc{\hbox to 3pc%
   {\bf  \vtop{\hfuzz=1pc\hsize=38pc\hyphenpenalty=10000\noindent{\bf 
  \the\sectnum.\the\subsectnum\ \ \ #1}}\hss}}%
			\hfill}%
   \nobreak%
			\vskip 1pc plus 4pt minus 2pt\nobreak\noindent\ignorespaces}%

\def\subsubsection#1{\goodbreak\vskip 1pc plus 4pt minus 2pt
   \hfuzz=3pc\leftskip=0pc\noindent\tenit #1 \nobreak\tenf\vskip 6pt plus 1pt
                                minus 1pt\nobreak\ignorespaces\leftskip=0pc}
%

\def\beginexample#1{\noindent\goodbreak\vskip 6pt plus 1pt minus 1pt
\noindent
  \hbox {\bf Example #1\hss}
  \nobreak\vskip 4pt plus 1pt minus 1pt \nobreak\noindent\ninef
  \global\advance
                \leftskip by\parindent\pn}
\def\endexample{\vskip 12pt\tenf\par
  \global\advance\leftskip by -\parindent
  }

\def\beginexercise#1{\noindent\goodbreak\vskip 6pt plus 1pt minus 1pt \noindent\global\normalbaselineskip=12pt
  \hbox {\bf Exercise #1\hss}
  \nobreak\vskip 4pt plus 1pt minus 1pt 
  \nobreak\noindent\ninef\global\advance\leftskip
                        by\parindent\pn}
\def\endexercise{\vskip 12pt\tenf\par
  \global\advance\leftskip by -\parindent
  }

\def\beginsection#1{\noindent\goodbreak\vskip 6pt plus 1pt minus 1pt \noindent\global\normalbaselineskip=12pt
  \hbox {\it #1\hss}
  \vskip 0.1pt plus 1pt minus 1pt \nobreak\noindent\ninef\global\advance
                \leftskip by\parindent\noindent\pn}
\def\endsection{\vskip 12pt\tenf\par
  \global\advance\leftskip by -\parindent
}

%


\def\proposition#1{\smskip\pn{\bf Proposition #1}\quad}
 
\def\definition#1{\smskip\pn{\bf Definition #1}\quad}

\def\ref{\smskip\pn}

\def\chapter#1#2{{\bf \centerline{\helbigbig
{#1}}}\bigskip\bigskip{\bf \centerline{\helbigbig
{#2}}}\bigskip\bigskip} 



\def\longpapertitle#1#2#3{{\bf \centerline{\helbigb
{#1}}}\bigskip{\bf \centerline{\helbigb
{#2}}}\bigskip\bigskip{\centerline{
by}}\bigskip{\bf \centerline{
{#3}}}\bigskip\bigskip} 


\def\nitem#1{\smskip\item{#1}}

\newcount\alphanum
\newcount\romnum

\def\alphaenumerate{\ifcase\alphanum \or (a)\or (b)\or (c)\or (d)\or (e)\or
(f)\or (g)\or (h)\or (i)\or (j)\or (k)\fi}
\def\romenumerate{\ifcase\romnum \or (i)\or (ii)\or (iii)\or (iv)\or (v)\or
(vi)\or (vii)\or (viii)\or (ix)\or (x)\or (xi)\fi}

\def\alist{\begingroup\vskip10pt\alphanum=1
\parskip=2pt\parindent=0pt \leftskip=3pc
\everypar{\llap{{\rm\alphaenumerate\hskip1em}}\advance\alphanum by1}}

\def\nolist{\begingroup\vskip10pt\alphanum=0
\parskip=2pt\parindent=0pt \leftskip=3pc
\everypar{\llap{\global\advance\alphanum by1(\the\alphanum)\hskip1em}}}

\def\romlist{\begingroup\vskip10pt\romnum=1
\parskip=2pt\parindent=0pt \leftskip=5pc
\everypar{\llap{{\rm\romenumerate\hskip1em}}\advance\romnum by1}}



\long\def\fig#1#2#3{\vbox{\vskip1pc\vskip#1
\prevdepth=12pt \baselineskip=12pt
\vskip1pc
\hbox to\hsize{\hfill\vtop{\hsize=25pc\noindent{\eightbf Figure #2\ }
{\eightpoint#3}}\hfill}}}

\long\def\widefig#1#2#3{\vbox{\vskip1pc\vskip#1
\prevdepth=12pt \baselineskip=12pt
\vskip1pc
\hbox to\hsize{\hfill\vtop{\hsize=28pc\noindent{\eightbf Figure #2\ }
{\eightpoint#3}}\hfill}}}

\long\def\table#1#2{\vbox{\vskip0.5pc
\prevdepth=12pt \baselineskip=12pt
\hbox to\hsize{\hfill\vtop{\hsize=25pc\noindent{\eightbf Table #1\ }
{\eightpoint#2}}\hfill}}}

 
\def\rightheadline#1{\headline{\tenrm\hfil #1}}


\long\def\leftfig#1#2{\vbox{\smskip\hsize=220pt
\vtop{{\noindent {\bf #1}}}
\smskip
\noindent
\vbox{{\noindent #2}}
}}

\long\def\rightfig#1#2#3{\vbox{\smskip\vskip#1
\prevdepth=12pt \baselineskip=12pt
\hsize=210pt
\smskip
\vbox{\noindent{\eightbold #2}
\hskip1em{\eightpoint#3}}
}}

\long\def\concept#1#2#3#4#5{\bigskip\hrule
\vbox{\hbox{\leftfig{#1}{#2} \hskip3em
\rightfig{#3}{#4}{#5}} \smskip}
\hrule\bigskip}


\long\def\bconcept#1#2#3#4#5#6#7{
\vbox{
\hbox to \hsize{\vtop{\par #1}}
\concept{#2}{#3}{#4}{#5}{#6}
\hbox to \hsize{\vtop{\par #7}}
\smskip}
}




\def\boxit#1{\vbox{\hrule\hbox{\vrule\kern3pt
                                \vbox{\kern3pt#1\kern3pt}\kern3pt\vrule}\hrule}}
\def\centerboxit#1{$$\vbox{\hrule\hbox{\vrule\kern3pt
                                \vbox{\kern3pt#1\kern3pt}\kern3pt\vrule}\hrule}$$}

\long\def\boxtext#1#2{$$\boxit{\vbox{\hsize #1\noindent\strut #2\strut}}$$}

%
%
%

\def\picture #1 by #2 (#3){
  \vbox to #2{
    \hrule width #1 height 0pt depth 0pt
    \vfill
    \special{picture #3} 
    }
  }

\def\scaledpicture #1 by #2 (#3 scaled #4){{
  \dimen0=#1 \dimen1=#2
  \divide\dimen0 by 1000 \multiply\dimen0 by #4
  \divide\dimen1 by 1000 \multiply\dimen1 by #4
  \picture \dimen0 by \dimen1 (#3 scaled #4)}
  }

%
%

\long\def\captfig#1#2#3#4#5{\vbox{\vskip1pc
\hbox to\hsize{\hfill{\picture #1 by #2 (#3)}\hfill}
\prevdepth=9pt \baselineskip=9pt
\vskip1pc
\hbox to\hsize{\hfill\vtop{\hsize=24pc\noindent{\eightbold Figure #4}
\hskip1em{\eightpoint#5}}\hfill}}}

%
%
%

\def\illustration #1 by #2 (#3){
  \vskip#2\hskip#1\special{illustration #3} 
    }

\def\scaledillustration #1 by #2 (#3 scaled #4){{
  \dimen0=#1 \dimen1=#2
  \divide\dimen0 by 1000 \multiply\dimen0 by #4
  \divide\dimen1 by 1000 \multiply\dimen1 by #4
  \illustration \dimen0 by \dimen1 (#3 scaled #4)}
  }


\newbox\graybox
\newdimen\xgrayspace
\newdimen\ygrayspace
%
%
%
%
%
%
%
%
%

\def\Textshade#1#2#3#4#5#6{%
    \xgrayspace=#4pt%
    \ygrayspace=#4pt%
    \def\grayshade{#3}%
    \def\linewidth{#5}%
    \def\theradius{#6}%
    \setbox\graybox=\hbox{\surroundboxa{#2}}%
    \hbox{%
    \hbox to 0pt{%
    \PScommands
}%
    \box\graybox}}%
%
%
\long%

\long%
\def\Parashade#1#2#3#4#5#6#7{%
    \xgrayspace=#4pt%
    \ygrayspace=#4pt%
    \def\grayshade{#3}%
    \def\linewidth{#5}%
    \def\theradius{#6}%
    \def\thevskip{#7pt}%
    \setbox\graybox=\hbox{\surroundboxb{#2}}%
    \vskip\thevskip%
    \hbox{%
    \hbox to 0pt{%
    \PScommands
}%
     \box\graybox}%
     \vskip\thevskip%
}%
%
%
%
\long\def\surroundboxa#1{\leavevmode\hbox{\vtop{%
\vbox{\kern\ygrayspace%
\hbox{\kern\xgrayspace#1%
      \kern\xgrayspace}}\kern\ygrayspace}}}
%
%
\long\def\surroundboxb#1{\leavevmode\hbox{\vtop{%
\vbox{\kern\ygrayspace%
\hbox{\kern\xgrayspace\vbox{\advance\hsize-2\xgrayspace#1}%
      \kern\xgrayspace}}\kern\ygrayspace}}}
%
%
%
\long\def\PScommands{%
\special{rawpostscript
/sharpbox{%
           newpath
           xmin ymin moveto
           xmin ymax lineto
           xmax ymax lineto
           xmax ymin lineto
           xmin ymin lineto
           closepath 
          }bind def
}%
\special{rawpostscript
/sharpboxnb{%
           newpath
           xmin ymin moveto
           xmin ymax lineto
           xmax ymax lineto
           xmax ymin lineto
          }bind def
}%
\special{rawpostscript
/sharpboxnt{%
           newpath
           xmin ymax moveto
           xmin ymin lineto
           xmax ymin lineto
           xmax ymax lineto
          }bind def
}%
\special{rawpostscript
/roundbox{%
           newpath
           xmin radius add ymin moveto
           xmax ymin xmax ymax radius arcto
           xmax ymax xmin ymax radius arcto
           xmin ymax xmin ymin radius arcto
           xmin ymin xmax ymin radius arcto 16 {pop} repeat
           closepath
          }bind def
}%
\special{rawpostscript
/sharpcorners{%
               sharpbox gsave grayshade setgray fill grestore 
               linewidth setlinewidth stroke
              }bind def
}%
\special{rawpostscript
/sharpcornersnt{%
               sharpboxnt gsave grayshade setgray fill grestore 
               linewidth setlinewidth stroke
              }bind def
}%
\special{rawpostscript
/sharpcornersnb{%
               sharpboxnb gsave grayshade setgray fill grestore 
               linewidth setlinewidth stroke
              }bind def
}%
\special{rawpostscript
/roundcorners{%
               roundbox gsave grayshade setgray fill grestore 
               linewidth setlinewidth stroke
              }bind def
}%
\special{rawpostscript
/plainbox{%
           sharpbox grayshade setgray fill 
          }bind def
}%
%
\special{rawpostscript
/roundnoframe{%
               roundbox grayshade setgray fill 
              }bind def
}%
\special{rawpostscript
/sharpnoframe{%
               sharpbox grayshade setgray fill 
              }bind def
}%
}%
%
%

\def\pshade#1{\Parashade{sharpcorners}{#1}{0.95}{10}{0.5}{10}{10}}


\def\boxit#1{\vbox{\hrule\hbox{\vrule\kern3pt
                                \vbox{\kern3pt#1\kern3pt}\kern3pt\vrule}\hrule}}

\def\boxitnb#1{\vbox{\hrule\hbox{\vrule\kern3pt
                                \vbox{\kern3pt#1\kern3pt}\kern3pt\vrule}}}

\def\boxitnt#1{\vbox{\hbox{\vrule\kern3pt
                                \vbox{\kern3pt#1\kern3pt}\kern3pt\vrule}\hrule}}

\def\boxitntnb#1{\vbox{\hbox{\vrule\kern3pt
                                \vbox{\kern3pt#1\kern3pt}\kern3pt\vrule}}}

\long\def\boxtext#1#2{$$\boxit{\vbox{\hsize #1\noindent\strut #2\strut}}$$}
\long\def\boxtextnb#1#2{$$\boxitnb{\vbox{\hsize #1\noindent\strut #2\strut}}$$}
\long\def\boxtextnt#1#2{$$\boxitnt{\vbox{\hsize #1\noindent\strut #2\strut}}$$}


\def\texshopbox#1{\boxtext{462pt}{\vskip-1.5pc\pshade{\vskip-1.0pc#1\vskip-2.0pc}}}
\def\texshopboxnt#1{\boxtextnt{462pt}{\vskip-1.5pc\pshade{\vskip-1.0pc#1\vskip-2.0pc}}}
\def\texshopboxnb#1{\boxtextnb{462pt}{\vskip-1.5pc\pshade{\vskip-1.0pc#1\vskip-2.0pc}}}


%
%
%
%
%
%
%
%
\font\helbigbig=cmr10 scaled 2500%
\font\helbigb=cmbx10 scaled 1500%
\font\eightbold=cmbx8%

\def\tenf{\hel}%
\def\tenit{\heli}%
\def\ninef{\ninehel}%
\def\nineit{\nineheli}%
%
%


\font\tenrm=cmr10%
\font\teni=cmmi10%
\font\tensy=cmsy10%
\font\tenbf=cmbx10%
\font\tentt=cmtt10%
\font\tenit=cmti10%
\font\tensl=cmsl10%

\def\tenpoint{\def\rm{\fam0\tenrm}%
\textfont0=\tenrm%
\textfont1=\teni%
\textfont2=\tensy%
\textfont\itfam=\tenit%
\textfont\slfam=\tensl%
\textfont\ttfam=\tentt%
\textfont\bffam=\tenbf%
\scriptfont0=\sevenrm%
\scriptfont1=\seveni%
\scriptfont2=\sevensy%
\scriptscriptfont0=\sixrm%
\scriptscriptfont1=\sixi%
\scriptscriptfont2=\sixsy%
\def\it{\fam\itfam\tenit}%
\def\tt{\fam\ttfam\tentt}%
\def\sl{\fam\slfam\tensl}%
\scriptfont\bffam=\sevenbf%
\scriptscriptfont\bffam=\sixbf%
\def\bf{\fam\bffam\tenbf}%
\normalbaselineskip=18pt%
\normalbaselines\rm}%

\font\ninerm=cmr9%
\font\ninebf=cmbx9%
\font\nineit=cmti9%
\font\ninesy=cmsy9%
\font\ninei=cmmi9%
\font\ninett=cmtt9%
\font\ninesl=cmsl9%

\def\ninepoint{\def\rm{\fam0\ninerm}%
\textfont0=\ninerm%
\textfont1=\ninei%
\textfont2=\ninesy%
\textfont\itfam=\nineit%
\textfont\slfam=\ninesl%
\textfont\ttfam=\ninett%
\textfont\bffam=\ninebf%
\scriptfont0=\sixrm%
\scriptfont1=\sixi%
\scriptfont2=\sixsy%
\def\it{\fam\itfam\nineit}%
\def\tt{\fam\ttfam\ninett}%
\def\sl{\fam\slfam\ninesl}%
\scriptfont\bffam=\sixbf%
\scriptscriptfont\bffam=\fivebf%
\def\bf{\fam\bffam\ninebf}%
\normalbaselineskip=16pt%
\normalbaselines\rm}%

\font\eightrm=cmr8%
\font\eighti=cmmi8%
\font\eightsy=cmsy8%
\font\eightbf=cmbx8%
\font\eighttt=cmtt8%
\font\eightit=cmti8%
\font\eightsl=cmsl8%

\def\eightpoint{\def\rm{\fam0\eightrm}%
\textfont0=\eightrm%
\textfont1=\eighti%
\textfont2=\eightsy%
\textfont\itfam=\eightit%
\textfont\slfam=\eightsl%
\textfont\ttfam=\eighttt%
\textfont\bffam=\eightbf%
\scriptfont0=\sixrm%
\scriptfont1=\sixi%
\scriptfont2=\sixsy%
\scriptscriptfont0=\fiverm%
\scriptscriptfont1=\fivei%
\scriptscriptfont2=\fivesy%
\def\it{\fam\itfam\eightit}%
\def\tt{\fam\ttfam\eighttt}%
\def\sl{\fam\slfam\eightsl}%
\scriptscriptfont\bffam=\fivebf%
\def\bf{\fam\bffam\eightbf}%
\normalbaselineskip=14pt%
\normalbaselines\rm}%

\font\sevenrm=cmr7%
\font\seveni=cmmi7%
\font\sevensy=cmsy7%
\font\sevenbf=cmbx7%

\def\sevenpoint{%
   \def\rm{\sevenrm}\def\bf{\sevenbf}%
   \def\smc{\sevensmc}\baselineskip=12pt\rm}%

\font\sixrm=cmr6%
\font\sixi=cmmi6%
\font\sixsy=cmsy6%
\font\sixbf=cmbx6%

\fontdimen13\tensy=2.6pt%
\fontdimen14\tensy=2.6pt%
\fontdimen15\tensy=2.6pt%
\fontdimen16\tensy=1.2pt%
\fontdimen17\tensy=1.2pt%
\fontdimen18\tensy=1.2pt%

\def\tenf{\tenpoint}%
\def\ninef{\ninepoint}%
%
